\documentclass[namedreferences]{solarphysics}
\usepackage{longtable}
\usepackage[optionalrh]{spr-sola-addons} 
\usepackage{epsfig}          
\usepackage{graphicx} 
\usepackage{rotating}
\usepackage{courier}         
\usepackage{amssymb}        
\usepackage{color}           
\usepackage{url}             
\usepackage{ulem}
\usepackage{float}
\usepackage{perpage}
\usepackage{here}

\usepackage{caption}

\newcommand{\be}{\begin{equation}}
\newcommand{\ee}{\end{equation}}
\newcommand{\beq}{\begin{eqnarray}}
\newcommand{\eeq}{\end{eqnarray}}

\newcommand{\annG}{   {\it Annales Geophysicae}}
\newcommand{\aap}{    {\it Astron. Astrophys.}}

\newcommand{\apj}{    {\it Astrophys. J.}}

\newcommand{\grl}{    {\it Geophys. Res. Lett.}}

\newcommand{\jgr}{    {\it J. Geophys. Res.}}

\newcommand{\solphys}{{\it Solar Phys.}}

\begin{document}

\normalem
\begin{article}
\begin{opening}

\title{Solar Wind Electron Strahls Associated with a
High-Latitude CME: \emph{Ulysses} Observations}
\author{M. \surname{Lazar}$^{1,2}$ $\sep$ J. \surname{Pomoell}$^{1}$ $\sep$
S. \surname{Poedts}$^{1}$ $\sep$ C. \surname{Dumitrache}$^{3}$
$\sep$ N.A. \surname{Popescu}$^{3}$}

\runningauthor{M. Lazar \emph{et al.}} \runningtitle{Solar Wind
Electron Strahls: Observations of a High-Latitude CME}

\institute{$^1$ Center for Plasma Astrophysics, K.U. Leuven,
Celestijnenlaan 200B, 3001 Leuven, Belgium \\ email:
\url{Marian.Lazar@wis.kuleuven.be};
\url{Jens.Pomoell@wis.kuleuven.be};
\url{Stefaan.Poedts@wis.kuleuven.be} \\
$^2$ Institut f\"ur Theoretische Physik, Lehrstuhl IV: Weltraum- und
Astrophysik, Ruhr-Universit\"at Bochum, D-44780 Bochum, Germany \\
email: \url{mlazar@tp4.rub.de}\\
$^3$ Astronomical Institute of Romanian Academy,040557 Bucharest,
Romania \\ email: \url{crisd@aira.astro.ro};
\url{nedelia@aira.astro.ro}}
\date{Received ; accepted }

\begin{abstract}
Counterstreaming beams of electrons are ubiquitous in coronal mass
ejections (CMEs) - although their existence is not unanimously
accepted as a necessary and/or sufficient signature of these events.
We continue the investigations of a high-latitude CME registered by
the \emph{Ulysses} spacecraft on January 18\,--\,19, 2002
(Dumitrache, Popescu, and Oncica, Solar Phys. {\bf 272}, 137, 2011),
by surveying the solar wind electron distributions associated with
this event. The temporal-evolution of the pitch-angle distributions
reveal populations of electrons distinguishable through their
anisotropy, with clear signatures of i)~electron strahls,
ii)~counter-streaming in the magnetic clouds and their precursors,
and iii)~unidirectional in the fast wind preceding the CME. The
analysis of the counter-streams inside the CME allows us to
elucidate the complexity of the magnetic-cloud structures embeded in
the CME and to refine the borders of the event. Identifying such
strahls in CMEs, which preserve properties of the low $\beta < 1$
coronal plasma, gives more support to the hypothesis that these
populations are remnants of the hot coronal electrons that escape
from the electrostatic potential of the Sun into the heliosphere.
\end{abstract}


\end{opening}

\section{Introduction}\label{in}

\emph{Ulysses} was launched in October 1990 to explore the
heliosphere from the solar Equator to the Poles. The observations
obtained enable a characterization of the particle populations
constituting the interplanetary plasma both out of the Ecliptic
plane as well as over a range of heliocentric distances exceeding 1
astronomical unit (AU). Opportunities have also been opened to
characterize the magnetic field structure and topology, and extract
more insights from their correlations with the velocity
distributions of charge particles injected by the Sun in the
heliosphere.


\subsection{Electron Strahls in the Solar Wind}

The electron-beaming strahl is well known up to 1~AU as a sharply
magnetic-field-aligned energetic population of electrons usually
moving anti-Sunward \cite{ro77,pi87a}, carrying with it a
significant amount of heat flux in the solar wind. Measurements by
the \emph{Helios} spacecraft revealed a high variability of the
energetic electron distributions with solar-wind properties and
different heliospheric distances from 0.3~AU up to 1~AU in the
Ecliptic (\citeauthor{pi87a} \citeyear{pi87a, pi87b, pi87c};
\citeauthor{an12} \citeyear{an12}). More specifically, in the
fast-wind the electron velocity distributions can be highly
anisotropic and skewed with respect to the magnetic-field direction,
showing a narrow strahl directed along the magnetic field away from
the Sun and occasionally exhibiting a second, less intense strahl
directed toward the Sun.

The variation of the electron strahl with heliocentric distance and
heliographic latitude has also been studied using observations
provided by \emph{Ulysses}, \emph{Wind}, and the Advanced
Composition Explorer (ACE) \cite{ha96, ma05, an12}. Recently,
indications for the existence of a suprathermal solar-wind strahl at
10~AU have been presented \cite{wa13}. The strahl becomes less
focused to the magnetic-field direction at larger heliocentric
distances \cite{pi87b,ha96}, and the angular width is broader than
it would be under adiabatic expansion from the corona \cite{le83}.
Moreover, the halo density is observed to increase at the expense of
the strahl density that declines with increasing heliocentric
distance, suggesting that the halo is formed as a result of
pitch-angle scattering of the strahl \cite{ma05}. Since Coulomb
collisions, which are not efficient at large heliocentric distances,
can be excluded fluctuations in the electric and magnetic fields
driven locally by the plasma anisotropies as well as the strahls
themselves are considered responsible for scattering the particles
\cite{vo05,pa07,pi11}.

\subsection{Electron Counter-streaming Strahls During CME Events}

Coronal mass ejections (CMEs) are huge clouds of magnetized plasma
originating from the Sun and propagating with velocities ranging
from several hundreds to several thousands of kilometers per second.
Although each event is unique, CMEs are commonly identified in the
solar-wind plasma by a set of characteristic features, such as an
intense magnetic-field [$B$] with a topology exhibiting a smooth
rotation and lower levels of fluctuations, low proton and electron
temperatures [$T$] (implying low plasma $\beta$ parameters,
\emph{i.e.}, $\beta \equiv 8\pi n k_{\rm B} T /B^2 < 1$, where $n$
denote the number density), as well as an enhanced abundance of
helium or heavy ions \cite{ri10}.

In addition, bi-directional electron beams are frequently observed
during CMEs \cite{sk00a, ni08, an12} and are interpreted as a
signature of closed magnetic-field lines with both foot points
connected to the solar corona \cite{go87}. These are energetic
electrons flowing in both directions, parallel (0$^{\rm o}$
pitch-angle) and antiparallel (180$^{\rm o}$ pitch-angle) to the
magnetic field. Although in the CMEs the electron velocity
distribution can show intense counter-streaming strahls that
characterize their internal flux-rope-like magnetic topology,
occasionally, counter-streams do not appear to be associated with
CMEs or may extend beyond the conventional CME intervals
\cite{ri10}. Counter-streaming electrons are also attributed to
interplanetary shocks, including the Earth's bow shock and
corotating interaction regions (CIRs) \cite{fe82, go93}, as well as
depletions of halo particles around 90$^{\rm o}$ pitch-angle
\cite{go02, sk06}.

From the early spaceborne missions, the observations of
unidirectional or counterstreaming strahls of electrons were only
available for relatively limited intervals in time and space, but
modern instruments provide better observations of these populations.
Recently, a list of counterstreaming events has been established,
but only for CMEs detected up to 1~AU in the Ecliptic \cite{an12}.
During a nine-month period in 1998, the ACE spacecraft
($\approx$~1~AU) detected 32 counter-streaming electron events (an
average of 3.65 per month) with durations ranging from 5 to 146
hours (a median length of 45 hours), and with a 60\,--\,100$\;\%$
presence during CMEs \cite{sk00a}. This rate is comparable to that
measured previously by the third International Sun-Earth Explorer
(ISEE-3) at times when the spacecraft was not magnetically connected
to the Earth's bow shock \cite{go87}. However, despite the vast
number of studies analyzing CMEs identified in the Ecliptic, only
few contain analyses of the electron counter-streaming variability
(\citeauthor{sk00a}, \citeyear{sk00a,sk00b}; \citeauthor{ni08},
\citeyear{ni08}; \citeauthor{an12}, \citeyear{an12}). On the other
hand, the observations of high-latitude CMEs are in general limited
to outlining their bulk properties (\citeauthor{eb09},
\citeyear{eb09}; \citeauthor{du11}, \citeyear{du11}), but evidence
of their extension at high latitudes in the heliosphere includes
indications of counterstreaming electrons \cite{re03}.

Magnetic clouds (MCs) are observed in a subset ($\approx 1/3$) of
CMEs and partially retain their characteristics as they propagate
(see \citeauthor{st11}, \citeyear{st11} and references therein), but
counter-streaming electrons are not necessarily present in the cloud
portion of the CME \cite{sk00b}. Thus, according to a survey (from
March 1998) of the same magnetic cloud at different radial distances
in the solar wind, counter-streaming electrons were detected by ACE
only in the noncloud portion of the surrounding CME, and
counter-streams observed by \emph{Ulysses} at 5~AU were present in
the cloud but only between the forward and reverse shocks. This
suggests that rather than being a signature of the CME, the
counterstreaming populations are shock-related, being obvious in
corotating shocks and forward-reverse shock pairs driven by the
over-expansion of CMEs that have a high internal pressure
\cite{go94b}.

\begin{figure}[h]
\includegraphics[width=100mm]{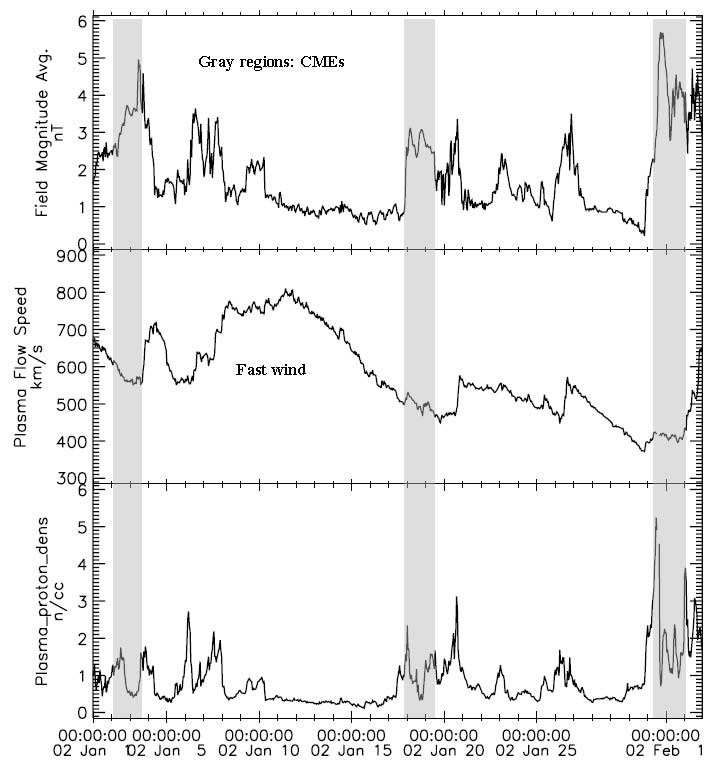}
    \caption{CDAWeb \emph{Ulysses} data plots of the magnetic field, plasma flow
    speed, and proton density for the interval DOY 01 00:00:00 to DOY 32
    24:00:00, year 2002. Shaded regions indicate three CMEs as reported
    by Ebert \emph{et al.} (2009).} \label{f1}
\end{figure}

\begin{sidewaysfigure}
\centering
\includegraphics[trim=1.6cm 0.0cm 4cm 0cm, clip, width=190mm]{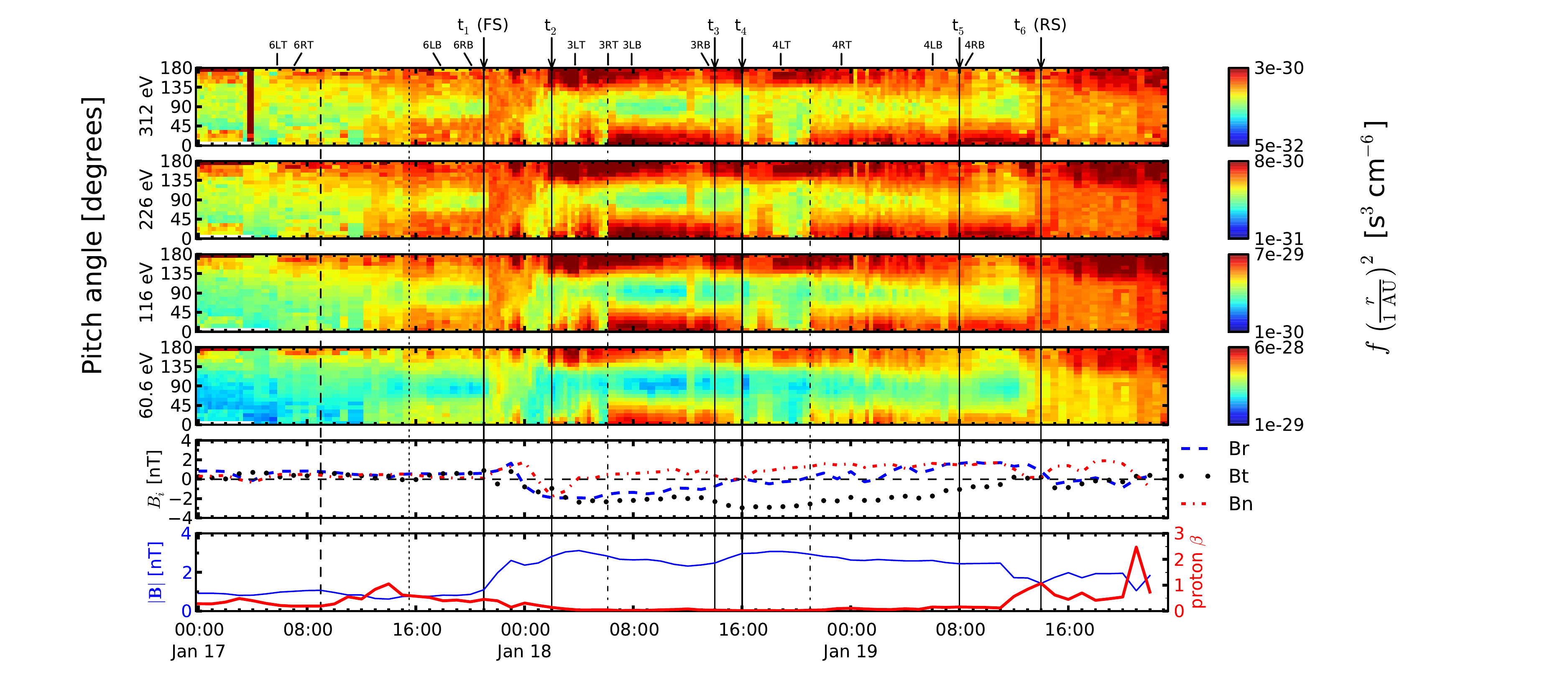}
\caption{Electron PADs as a function of time for four energy
channels: 60.0, 116, 226, and 312 eV (top). Bottom panels display
the magnetic-field components [$B_i$], magnitude [$B$], and proton
plasma $\beta$. Six designated times [$t_1$ to $t_6$] discussed in
the text have been indicated with arrows. The short solid lines
indicate times for which additional data have been plotted, the
number giving the figure number while the letter combination
indicating the corresponding panel (L = Left panel, R = Right panel,
T = Top panel, B = Bottom panel).} \label{f2}
\end{sidewaysfigure}

\subsection{Goal of This Study}

The present work is focused on a systematic analysis of the electron
velocity distributions measured by the Solar Wind Observations Over
the Poles of the Sun (SWOOPS) Electron instrument onboard {\it
Ulysses} during a high-latitude CME observed on January 18\,--\,19,
2002, when the {\it Ulysses} spacecraft was situated at 2.7~AU
distance from the Sun and at 62$^{\rm o}$ North heliographic
latitude relative to the Ecliptic. This CME is included in the
\emph{Ulysses} CME list (\url{swoops.lanl.gov/cme_list.html}), and
recently this event was tracked back to its solar origin by
\inlinecite{du11}, hereafter called Article~I. Article~I examined
the correlation between the magnetic field, velocity, temperature,
and density, to enable the identification of the CME and its
components such as the forward shock, magnetic clouds, and the
reverse shock. Here, we analyze the electron velocity distributions
and their temporal evolution, giving particular attention to
suprathermal populations that show enhanced field-aligned fluxes of
electrons. Our main purpose is to identify the intervals of
bi-directional (counter-streaming) or uni-directional electrons, and
correlate them with the CME profiles found in Article~I. Moreover,
the analysis is extended to the ambient solar wind before and after
the CME event, allowing us to outline the features of the
counter-streaming strahls that are specific to the CMEs.

Because CMEs reproduce and preserve coronal-plasma states, the
enhanced strahls present in CMEs support the hypothesis that these
populations are remnants of the hot coronal electrons, which escape
from the electrostatic potential of the Sun into space without
suffering collisions and conserving their magnetic moments
\cite{ma05}. Therefore, a study of their properties can give hints
of the prevailing conditions in the solar corona.

\section{Data Selection and Analysis}

In order to set the context of the CME event examined in this
article, Figure~\ref{f1} presents the temporal evolution of the
magnetic-field magnitude, plasma flow speed, and proton number
density measured by \emph{Ulysses} during the first 33 days of 2012.
The shaded regions indicate three CMEs in this interval \cite{eb09},
with our event identified (middlemost) at the day-of-year (DOY)
18-19, and preceded by a long-lasting fast-wind interval. Here the
analysis is focused on a period of 12 days, between DOY 10 and DOY
23, which includes the CME and intervals of ambient solar wind
before and after the ejecta.

The present study is based on the \emph{Ulysses}/SWOOPS electron
pitch-angle data published by the ESA-RSSD Web service \cite{sk12}.
These data are in the solar-wind frame, and have been corrected for
the spacecraft potential \cite{sk12, sc94}. Temporal evolutions of
the electron pitch-angle distributions (PADs), as in
Figure~\ref{f2}, are used to depict the intervals with uni- or
bi-directional strahls. The phase-space distributions are
individually shown for each energy channel (Table \ref{tab1}), and
are normalized to 1~AU distance by multiplying the data by the
square of heliocentric distance measured in units of AU. This
normalization enables a consistent color scale to be used for plots
from all spacecraft positions. The strahls appear as peaks of
intense fluxes (densities) at 0$^{\rm o}$ or/and 180$^{\rm o}$
pitch-angle, and are more prominent at energies above the break
between the core (Maxwellian fit) and high-energy tails. Thermal
populations are, therefore, those with kinetic energy below the
break point, while suprathermal energies are always above the break
point. In the solar wind in the absence of CMEs and other
disturbances, the break point is around 70~eV at 1~AU, and drops
with increasing radial distance from the Sun, with a typical value
of 30~eV near 5~AU.

Additional analysis of the magnetic field components, magnitude, and
the proton plasma $\beta$ (bottom panels in Figure \ref{f2}) enable
us to correlate magnetic-field topology with the presence of
strahls, and provide their orientation. It is worthwhile noting that
beaming particles at 0$^{\rm o}$ pitch-angles are traveling parallel
to the magnetic field and in the same direction [$+{\bf B}$], which
in general can be either toward [$B_r < 0$] or away [$B_r > 0$] from
the Sun. When \emph{Ulysses} flew over the North Pole of the Sun in
January 2002, the heliospheric magnetic field was registered mostly
pointing towards the Sun \cite{jo03}. In such a case, a
uni-directional strahl peaked at 180$^{\rm o}$ pitch-angle indicates
an anti-parallel flow directed away from the Sun. Here we call
\emph{parallel} the strahls centered at 0$^{\rm o}$ pitch-angle, and
\emph{anti-parallel} the strahls occurring at 180$^{\rm o}$
pitch-angle.

\begin{table}[h!]
\caption{Energy channels of the \emph{Ulysses}/SWOOPS electron
instrument and the corresponding electron speeds} \label{tab1}
\begin{tabular}{c c c c c c c c c c c } \hline
$E$ [eV] & 1.69 & 2.35 & 3.25 & 4.51 & 6.26& 8.65 & 12.1 & 16.8 & 23.2 & 31.9 \\
$v [10^3$km s$^{-1}$] & 0.77 & 0.90 & 1.06 & 1.25 & 1.48 & 1.74 &
2.05 & 2.42 & 2.84 & 3.33 \\ \hline
& 43.9 & 60.6 & 84.0 & 116 & 163 & 226 & 312 & 429 & 591 & 815 \\
& 3.91 & 4.59 & 5.41 & 6.35 & 7.53 & 8.87 & 10.42 & 12.22 & 14.34 & 16.84 \\
\end{tabular}
\end{table}

A single (uni-directional) peak of the PADs is a clear signature of
a strahl, but two opposite peaks, in directions parallel and
anti-parallel to the magnetic field, can either be an indication of
counter-streams, or can result from a thermal anisotropy of the
distribution, \emph{i.e.} an excess of parallel temperature
$T_\parallel > T_\perp$, where $\parallel$ and $\perp$ denote
directions parallel and perpendicular to the magnetic field,
respectively. We make the distinction starting from the premise that
two opposite peaks must show the same, constant symmetry at
different energy channels to indicate an anisotropy of the parallel
temperature. Otherwise, if these peaks are signatures of strahls,
their symmetry is expected to break down at some energy levels,
based on the fact that the strahls have different paths and origins,
\emph{e.g.} distinct foot points, or the fast wind. To objectively
assess the presence and extent of the strahls, and whether these are
uni- or bi-directional, we analyze snapshots of the PAD profiles at
any instant of interest and for each energy channel, ranging from $E
= 1.69\;$eV to $814\;$eV (see Table~\ref{tab1}).

\begin{sidewaysfigure}
\centering
\includegraphics[trim=1.5cm 0.0cm 0.5cm 0cm, clip, width=190mm]{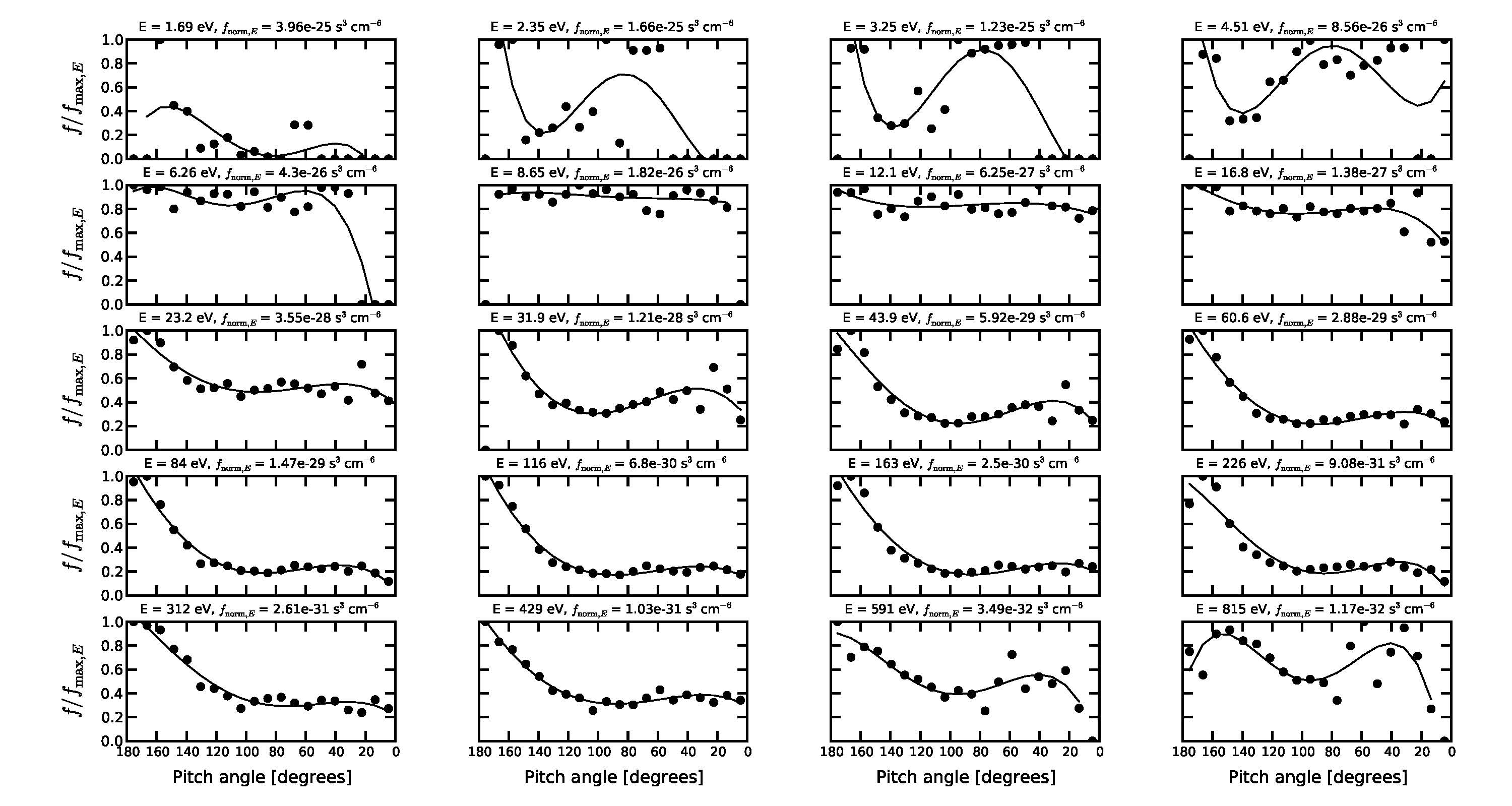}
\caption{PADs for all available energy channels on DOY 18 00:49:25.
Dots represent measured data of phase-space distributions [$f$]
normalized to the maximum value [$f_{{\rm max},E}$], and curves are
polynomial fits to the data.} \label{fb1}
\end{sidewaysfigure}

\begin{sidewaysfigure}
\centering
\includegraphics[trim=1.5cm 0.0cm 0.5cm 0cm, clip, width=190mm]{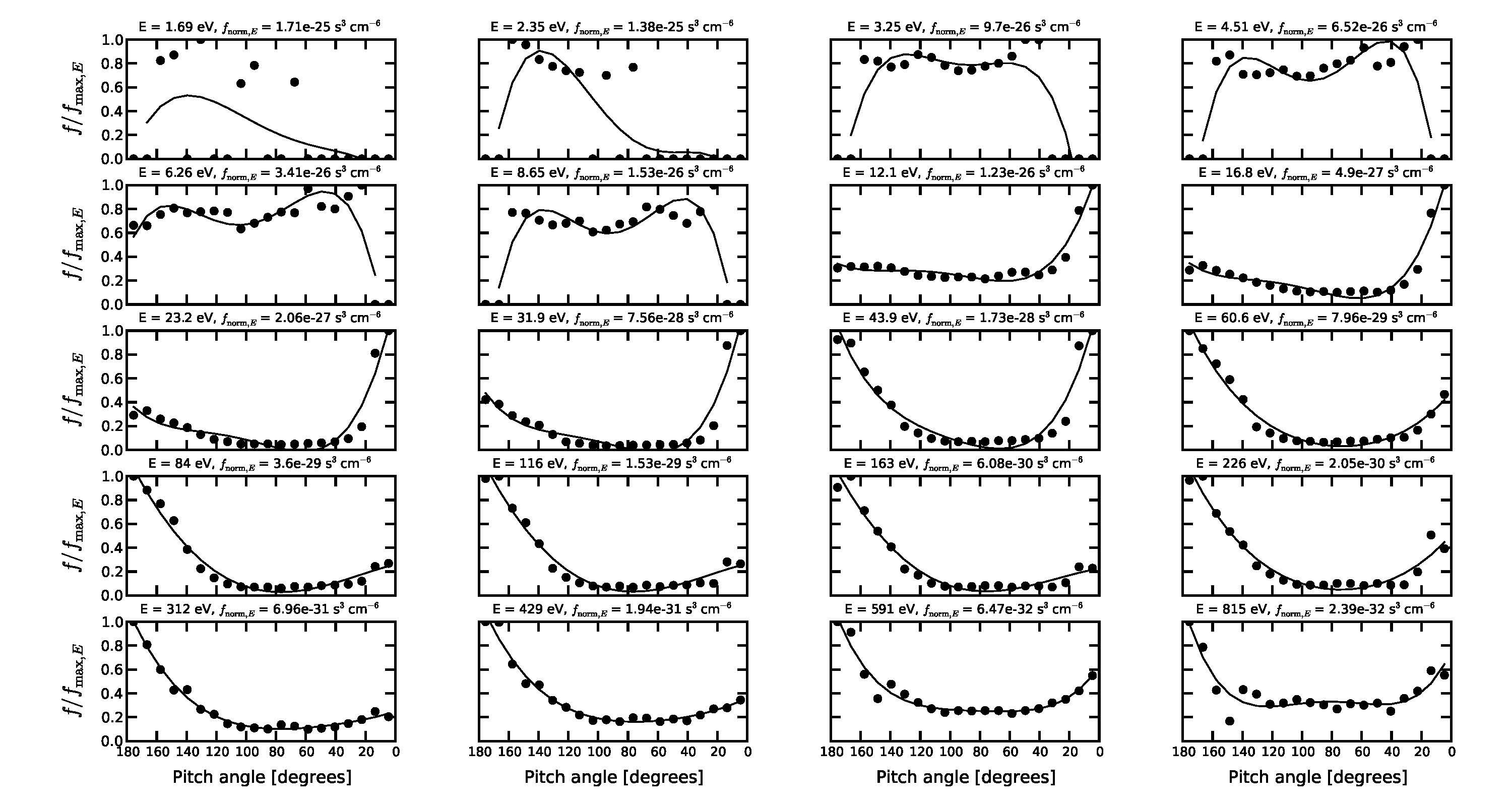}
\caption{PADs for all available energy channels (see Fig. \ref{fb1})
on DOY 18 01:58:43. The first signatures of counter-streams during
the CME event.} \label{fb2}
\end{sidewaysfigure}

To take a snapshot, we plot the PADs for all energy channels, like
the ones shown in Figures~\ref{fb1}, and \ref{fb2}, or for a single
(suprathermal) energy channel, as in Figures~\ref{f6}, and \ref{f7}
(bottom panels).
The strahls are manifested by the asymmetric peaks that occur at
0$^{\rm o}$ or/and 180$^{\rm o}$ pitch-angle. The parallel and
antiparallel peaks are compared by visual inspection, and to
highlight the peaks and depletions, the data is fitted with a
fourth-order polynomial that only serves as a guide for the eye and
omits the missing (0 level) data if present. Such a fitting model
can be used to quantify the main properties of the field-aligned
strahls, \emph{i.e.} the angular width and intensity \cite{pa07,
an12}. Traditionally, the method involves a Maxwellian, although
such a model describes well only the low-energy (core) populations
and introduces uncertainties and overestimates for suprathermal
populations. The implementation of a new power-law model is
presently under our consideration, since power-laws provide a good
fit for the high-energy tails of suprathermal distributions,
including the strahls \cite{la12}. Such an approach will enable us
to quantify their properties, but it is outside the scope of the
present study.

In order to build a complete picture for the full set of energy
channels, the pitch-angle data [$E$, $\theta$] is transformed into
the velocity space ($v_\parallel = v \cos \theta$, $v_\perp = v \sin
\theta$), where $v = (2E/m_e)^{0.5}$, with the corresponding values
given in Table~\ref{tab1}. This allows us to plot iso-contours of
the phase-space density in velocity space ($v_\parallel, v_\perp$),
{\it e.g.}, Figures~\ref{f6} and \ref{f7} (top panels), and
visualize the strahls as a (strong) bulging of the contour lines in
either the negative or positive $v_\parallel$-direction
(unidirectional strahls), or in both directions (counter-streams).
The positive $v_\parallel$-axis is directed along $+{\bf
B}$-direction of the mean magnetic field, and the coordinate system
($v_{\parallel}, v_{\perp}$) is centered on the bulk velocity of the
solar wind, which is nearly at the maximum of the distribution
function. Note that these velocity distributions are all gyrotropic
according to the method of construction from the measured data
\cite{sk12}.

Our study of the PADs is complemented by an analysis of the
electron-plasma moments, such as density and temperature, provided
by the SWOOPS team and obtained by a numerical integration of the
velocity-weighted electron distributions. To distinguish between the
intervals and distributions marked by the presence of strahls, we
compare their moments calculated either for the whole distribution,
{\it i.e.} total density [$n_{\rm total}$], and total temperature
[$T_{\rm total}$], or only for the components of the electron
distribution, below and above the core--halo energy break point,
{\it i.e.} [$n_{\rm c}$, $T_{\rm c}$] for the core, and [$n_{\rm
h}$, $T_{\rm h}$] for the halo.

\section{The January 18\,--\,19, 2002 CME Event}

Figure~\ref{f2} displays the electron pitch-angle distributions
(PADs) and their variability  with time and energy during the CME
event, {\it i.e.} for Days 17, 18, and 19 of 2002. Out of the
available energy channels, ranging from $1.69\;$eV to $814\;$eV (see
Appendix~A), we display four suprathermal energies, 60.6 (bottom),
116, 226, and $312\;$eV (top), which are relevant for the strahl
component. The color code runs from blue (low values) to red (high
values) and is different for each energy channel. Six solid bars
mark the principal structural elements of the CME identified in
Article~I. A shock front occurs at $t_1=17.875$ (DOY 17 21:00:00).
The CME event contains two MCs, each corresponding to a well-shaped
peak of the magnetic-field strength in Figure~1, top panel. The
first MC, denoted MC1, lasts from $t_2=18.083$ (DOY 18 01:59:31) to
$t_3=18.583$ (DOY 18 13:59:31), and the second, denoted MC2, from
$t_4 = 18.667$ (DOY 18 16:00:28) until $t_5 = 19.333$ (DOY 19
07:59:31). A reverse shock is identified at $t_6 = 19.583$ (DOY 19
13:59:31). In the next subsections, these boundaries are contrasted
with those indicated by the temporal variation of the electron PADs,
and by the presence of uni- and bi-directional strahls.

\subsection{Inside the CME}

According to Figure~\ref{f2}, before the forward shock the magnetic
field was observed pointing outward from the Sun [$B_{r} >0$], but
this orientation changes sharply in the sheath after the shock. The
radial and tangential components change sign [$B_{r,t} <0$] at DOY
18 00:00:00, and the normal component does the same [$B_n < 0$]
after only one hour. Immediately after the shock the PAD appears
less anisotropic, without evidence of peaks. The first indications
of an anti-parallel peak occur shortly after the magnetic-field
reversal (DOY 18 00:00:00). Figure~\ref{fb1} presents the PADs for
all available energy channels on DOY 18 00:49:25, when an
anti-parallel intense peak is already visible at all energy channels
between 23.2 and 591 eV. Likewise, a second parallel peak arises
less regularly, as an indication that \emph{Ulysses} is approaching
the sector of counter-streaming electrons trapped in the CME
magnetic field. The parallel peak is less intense and less regularly
observed, probably because of beam-plasma instabilities (reminiscent
after the forward shock), which may interrupt the large scale
connection of the magnetic field lines with both footpoints to the
Sun.

\begin{figure}[h!]
\centering
\includegraphics[trim=1cm 1cm 2cm 1cm, clip,width=90mm]{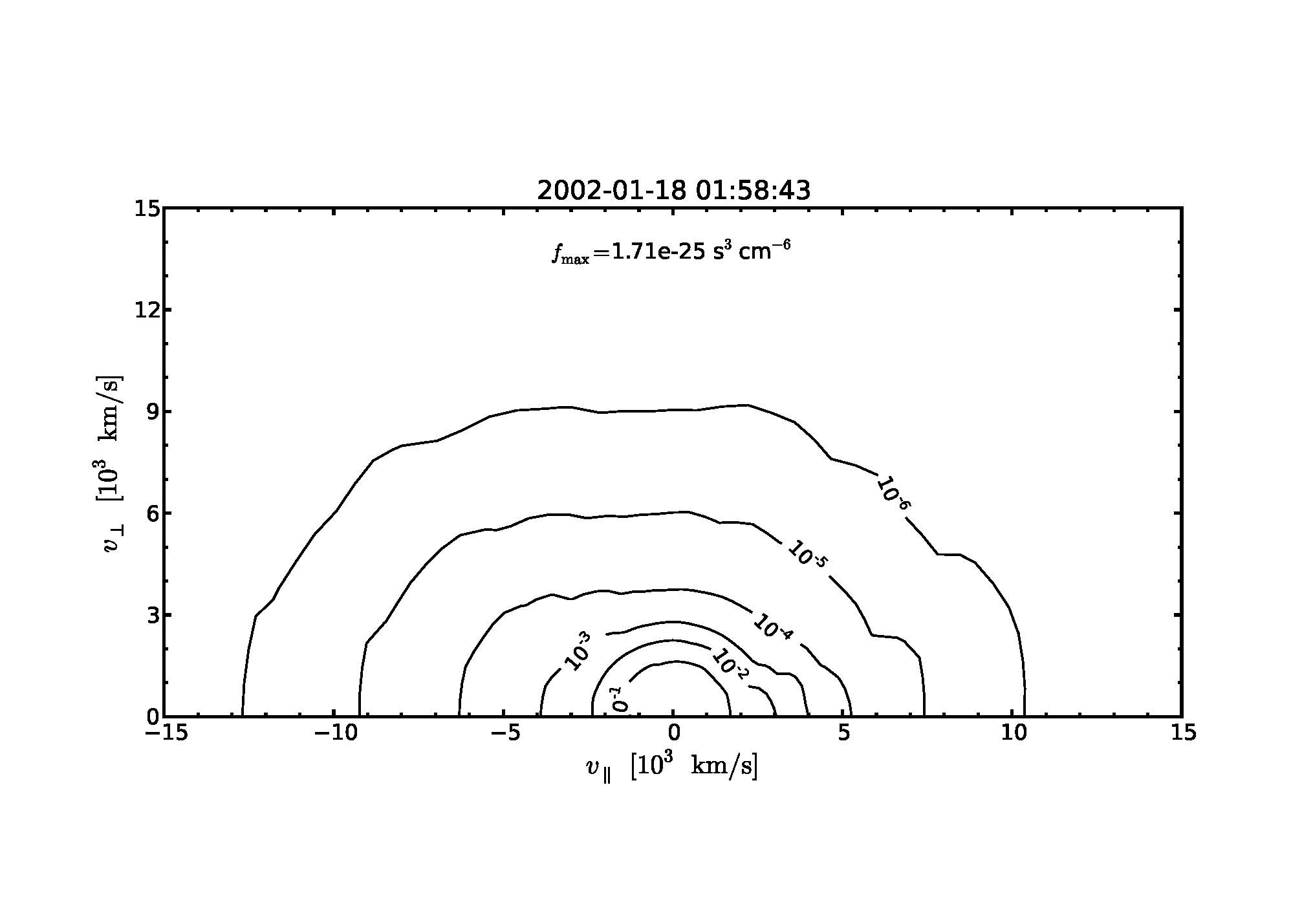} \hspace{0.5cm}
\includegraphics[trim=0 0 0.5cm 0, clip,width=90mm]{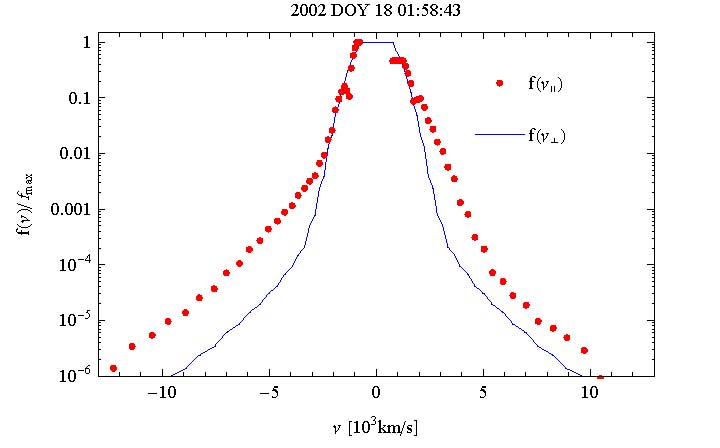}
\caption{Isocontours in velocity-space (top panel) and the parallel
and perpendicular cuts (bottom panel) measured on DOY 18 01:58:43}.
\label{fc1}
\end{figure}

Figure~\ref{fb2} presents the PADs with the first distinct
signatures of both the parallel and anti-parallel peaks on DOY 18
01:58:43. This is nearly the start time of the first magnetic cloud
MC1 identified in Article~I at $t_2=$ DOY 18 01:59:31. It is
important to emphasize the differences observed between the parallel
and anti-parallel peaks, because these differences, if they exist,
support the hypothesis that the two opposite peaks are signatures of
two counter-streaming strahls, and not indications of an excess of
parallel temperature [$T_\parallel > T_\perp$]. Based on this
hypothesis, the presence of strahls is confirmed by a systematic
analysis of the PADs for all the energy channels, and all the events
from an interval of interest. Such a contrast between the opposite
strahls is evident in Figure~\ref{fb2}, where the parallel peak is
visible at lower energies, starting at 12.1 eV, while the
anti-parallel peak arises at higher energies, $\approx$ 23.2 eV.
These two peaks become comparable at 43.9 eV, whereas at higher
energies, as the contribution from suprathermal populations
increases, the anti-parallel peak becomes more pronounced.

\begin{sidewaysfigure}
\centering
\includegraphics[trim=1.5cm 0.0cm 0.5cm 0cm, clip, width=190mm]{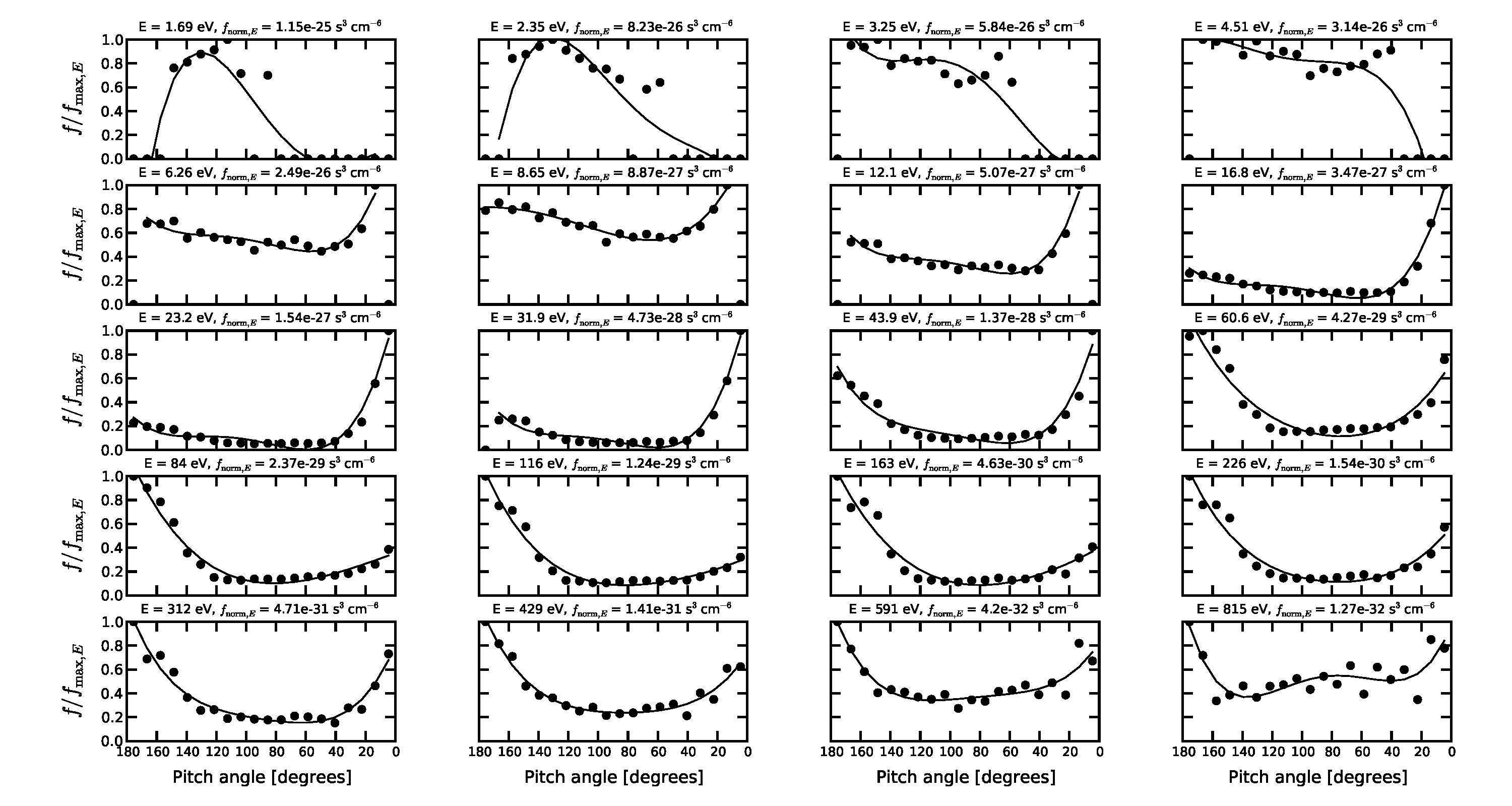}
\caption{PADs for all available energy channels (see Fig. \ref{fb1})
on DOY 18 02:16:02.} \label{fb3}
\end{sidewaysfigure}

For the same event (DOY 18 01:58:43), two bulging lobes of the
isocontour lines are also distinguishable in Figure~\ref{fc1} (top
panel), indicating an anti-parallel strahl more intense (at least
one order of magnitude) than the parallel strahl (compare, for
instance, higher suprathermal energies). Moreover, the parallel cut
in the velocity distribution shown in the bottom panel of
Figure~\ref{fc1} (filled circles) confirms the existence of these
two peaks, the parallel one being indeed close to $v_\parallel =
2.05 \times 10^3$ km s$^{-1}$ $\approx 12.1$ eV, but, contrary to
the indications in Figure~\ref{fb2}, the anti-parallel peak appears
at even lower energies, \emph{i.e.} $v_\parallel = 1.48 \times 10^3$
km s$^{-1}$ $\approx 6.26$ eV. At this energy, the map of PADs
indicates only an incipient tendency of enhancing the electron flux
in parallel direction (hidden, eventually, by a (weak)
thermalization, \emph{i.e.} $T_\parallel \gtrsim T_\perp$). This
tendency is confirmed by the temporal evolution of the PAD, see, for
instance, Figure~\ref{fb3}, which presents a snapshot on DOY 18
02:16:02, when the parallel peak appears at lower energies down to
$6.26$ eV. Based on these arguments, we can claim that the parallel
and anti-parallel peaks in Figures~\ref{fb2} (DOY 18 01:58:43
$\approx t_2$) mark the first counter-streaming signatures after the
forward shock, in MC1. A further inspection of the energy maps
reveals even more differences between these counter-streams, namely,
different intensities and angular widths, which also induce the idea
that parallel and anti-parallel strahls must have different origins.

\begin{figure}[h!]
\includegraphics[trim=1.5cm 1.5cm 2cm 1.5cm, clip, width=120mm]{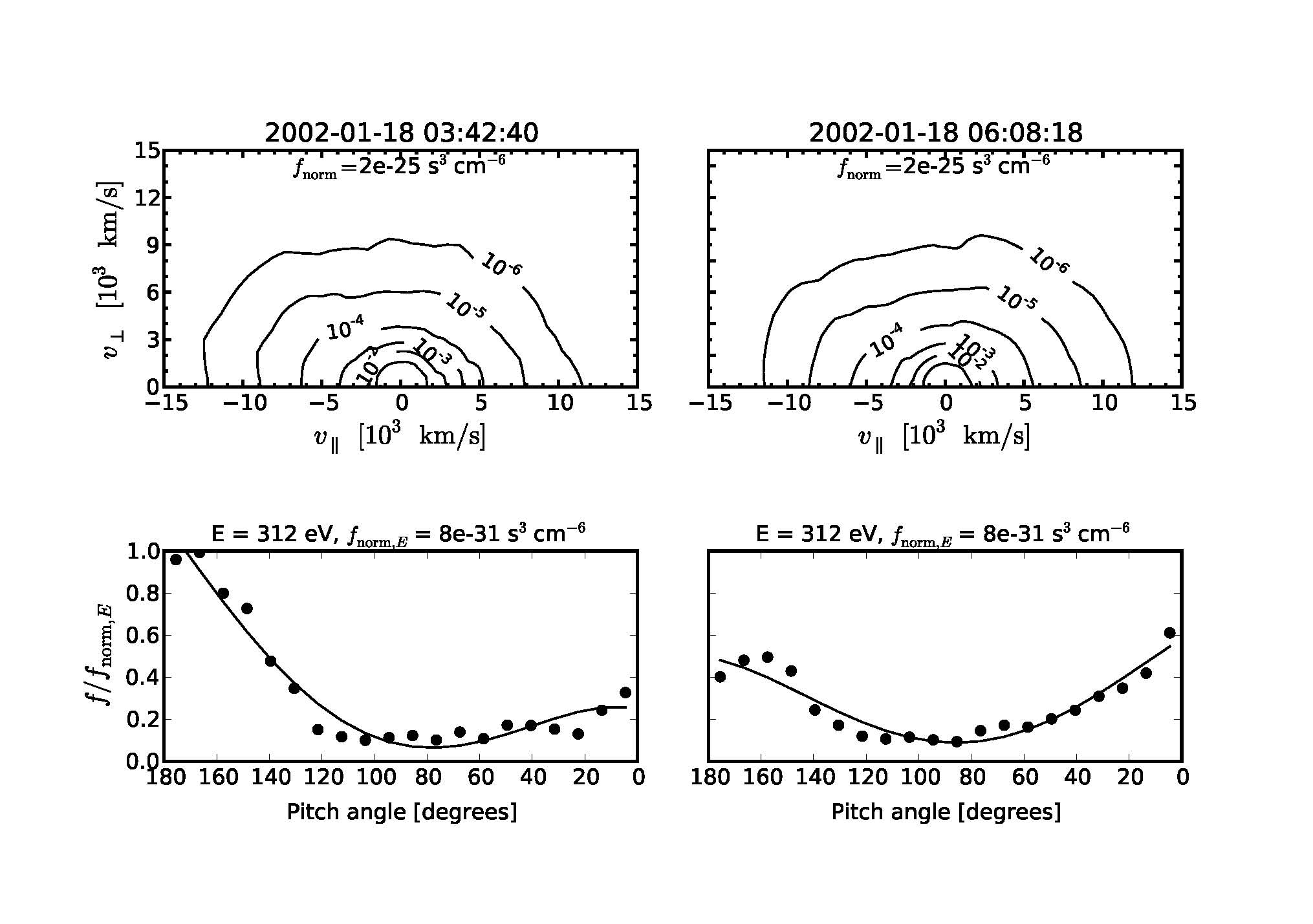}\\
\includegraphics[trim=1.5cm 1.5cm 2cm 1.5cm, clip, width=120mm]{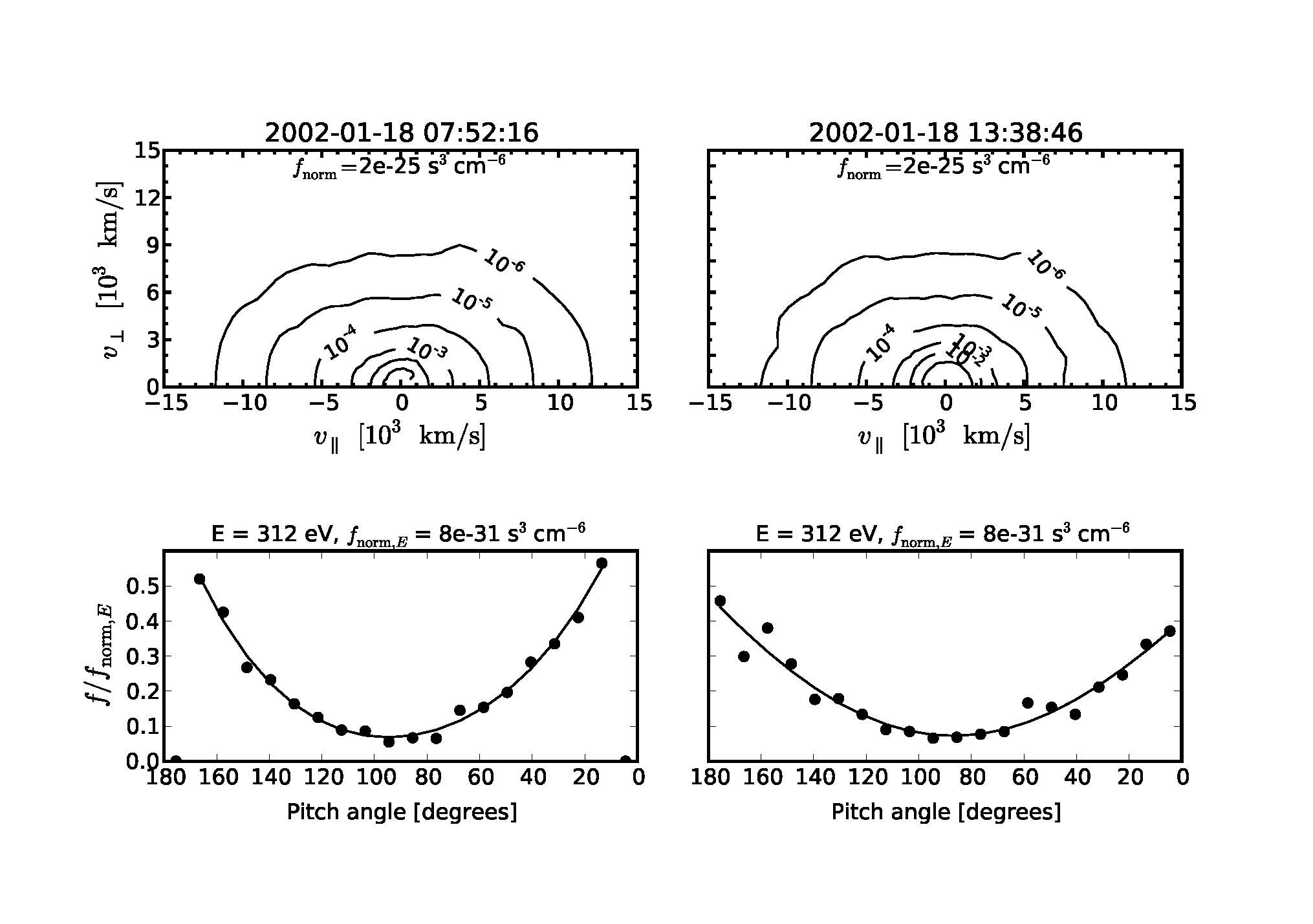}
    \caption{Snapshots of velocity distributions (contours) and PADs ($312\;$eV) from MC1.} \label{f6}
\end{figure}

\begin{figure}[h!]
\includegraphics[trim=1.5cm 1.5cm 2cm 1.5cm, clip, width=120mm]{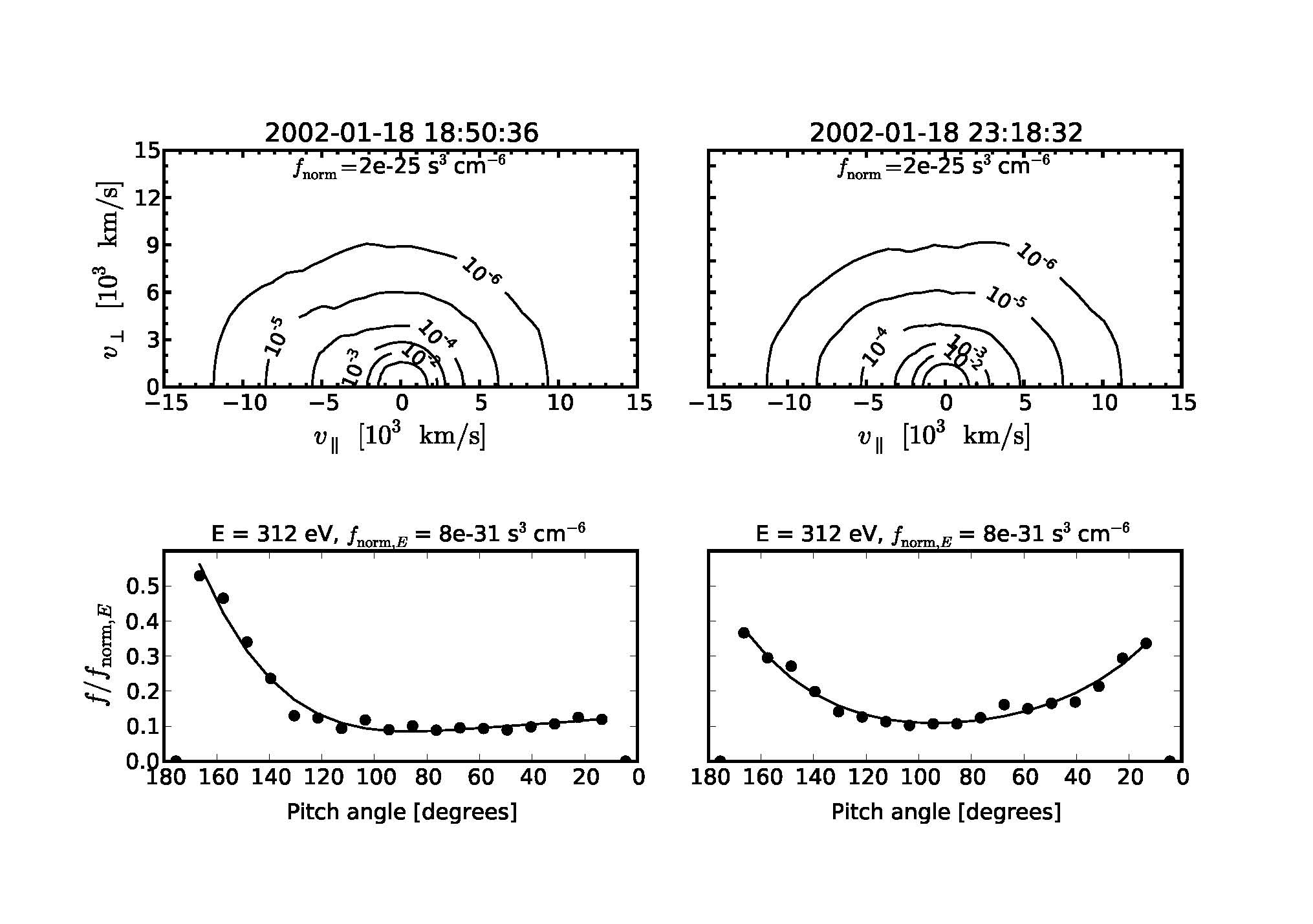}\\
\includegraphics[trim=1.5cm 1.5cm 2cm 1.5cm, clip, width=120mm]{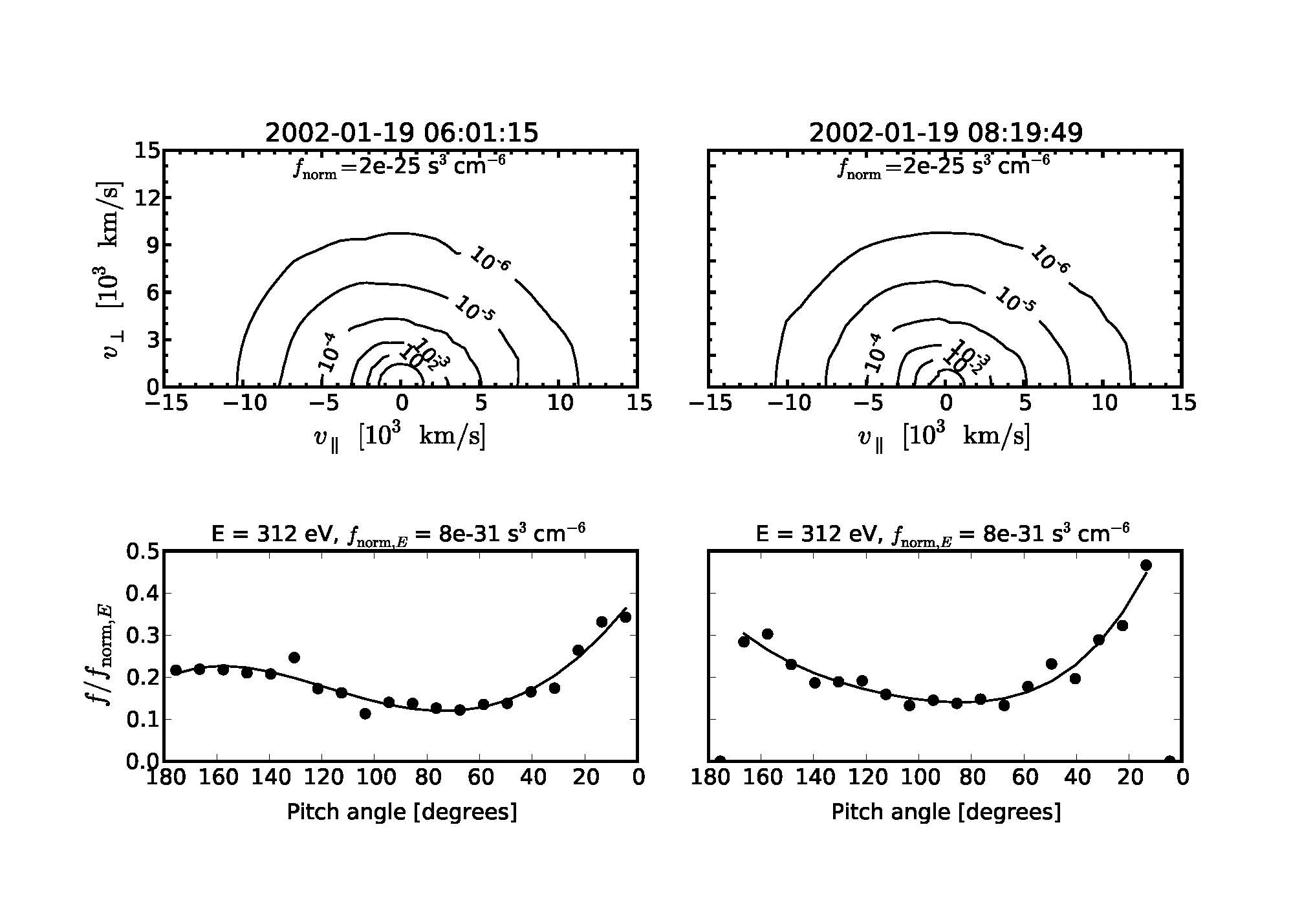}
    \caption{Snapshots of velocity distributions (contours) and PADs ($312\;$eV) from MC2.} \label{f7}
\end{figure}

At later times, \emph{i.e.} $t \geqslant t_2' =$ DOY 18 06:07:00
(first dash-dotted line in Figure~\ref{f2}), the counter-streaming
patterns are continuous, and more skewed and symmetric at all
suprathermal energies. Our analysis will therefore be limited to
follow their evolution at a single channel of higher energy,
\emph{e.g.} $E=$312 eV, which corresponds to an electron speed of
$\sim 10\times10^3\;$km s$^{-1}$ (see Table~\ref{tab1}). Four
samples relevant for the MC1 interval are displayed in
Figure~\ref{f6} with isocontours of the velocity distributions (top
panels) and the corresponding PADs (bottom panels). The contours of
the phase space density are logarithmically spaced and correspond to
fractions 10$^{-1}$, 10$^{-2}$, ..., 10$^{-6}$ of the normalization
value (provided in each panel). The strahls appear bending the round
shape of the isolines and are confirmed by the peaks in the bottom
panels. While the first case in the top--left panels is
representative for highly asymmetric counter-streams with a more
intense anti-parallel strahl, at the entrance of MC1, in the other
three events the counterstreams appear more balanced and symmetric.
For instance, the bottom--left panels display intense and symmetric
countermoving streams. The intensities of the strahls can easily be
compared as we have chosen the same normalization value for the
distribution functions, but a quantitative analysis will be
presented in a subsequent work. The parallel strahl is enhanced in
intensity, apparently at the expense of the first anti-parallel
strahl, but a complete reversal of the uni-directional strahl is not
observed, probably because of an incomplete rotation of the magnetic
field in MC1, see Figure~2, where $B_r < 0$ decreases in magnitude
but does not change sign in this interval. However, an additional
complete rotation (from negative to positive values) of the normal
component $B_n$ at the beginning of MC1, may be the origin of
intense but irregular parallel peaks observed until 07:00:00 (DOY
18) along with some remissions in intensity of the anti-parallel
peak.

Although perturbations appear at DOY 18 12:00:00, the
counter-streaming configurations in MC1 remain clear until DOY 18
16:00:00, simultaneous with both the radial and normal components of
the magnetic field vanishing. However, the analysis in Article~I
found that \emph{Ulysses} enters the second MC (MC2) at this time,
\emph{i.e.} $t_4 = 18.667 =$ DOY 18 16:00:28. Due to irregularities
of the parallel strahl, counter-streams become asymmetric again for
an interval of five hours following $t_4$ until about $t_4'=$
21:00:00 (second dash--dotted line in Figure~\ref{f2}). This
interval therefore appears similar to the interval of four hours
$[t_2, t_2']$ at the beginning of MC1. In addition, the radial
component of the magnetic field changes sign again, pointing toward
the Sun ($B_r < 0$) after $t_4$. These similarities between the
intervals $[t_4,t_4']$  and $[t_2, t_2']$ give further support to
the hypothesis that the CME event contains two distinct MCs.
However, the end of MC1 indicated in Article~I to be at $t_3=18.583$
(DOY 18 13:59:31) is only vaguely reflected in the profile of the
PADs, which does not show a significant change, but only a slight
remission of intensity of the parallel strahl. As discussed above,
after $t_4 = 18.667$, this beam is markedly diminished or seems to
be completely suppressed for an interval of approximately 1.5 hours
(between 18:00:00 and 20:00:00). Such a profile of PADs with a
single strahl in the antiparallel direction is shown in
Figure~\ref{f7}, top--left panels. Short periods of uni-directional
strahls inside the CME intervals could indicate disconnected or open
magnetic-field lines, interweaved with the closed field lines
\cite{go95,la97,sk00a}. Suppression of the parallel flow can be
observed at all the energetic channels reliable for suprathermal
data.

According to Article~I, the second magnetic cloud MC2 lasts until
$t_5 = 19.333$ (DOY 19 07:59:31), and is followed by a reverse shock
at $t_6 = 19.583$ (DOY 19 13:59:31). Four samples of anisotropic
velocity distributions and PADs from this interval are shown in
Figure~\ref{f7}. The temporal evolution from the top--left panels to
the bottom--right panels indicates a reversal of the uni-directional
strahl from the anti-parallel to parallel direction, respectively.
This is well correlated with the rotation of the radial component
$B_r$ of magnetic field, from negative to positive values
(Figure~\ref{f2}). Counter-streaming strahls are in general
associated with these reversals in their intermediary phase, but not
always. Similar reversals of uni-directional strahls have been
reported from the \emph{in-situ} observations by the Solar
TErrestrial RElations Observatory (STEREO) spacecraft A \cite{ro09},
with the difference that suprathermal electrons were not
counter-streaming, suggesting that the magnetic-field lines forming
the flux rope were open. Here the existence of counter-streams of
suprathermal electrons is evident in intermediary phases. For
instance, Figure~\ref{f7} indicates a gradual reversing of the
magnetic field (closed field lines). Although the radial
magnetic-field component (Figure~\ref{f2}) does not exhibit a smooth
variation during MC2, undergoing several short-time changes of sign
(gradient reversals), the bipolar signature is evident from the
larger scale trend in the entire interval of this cloud.

\begin{figure}[h]
\begin{center}
\includegraphics[width=0.5\textwidth]{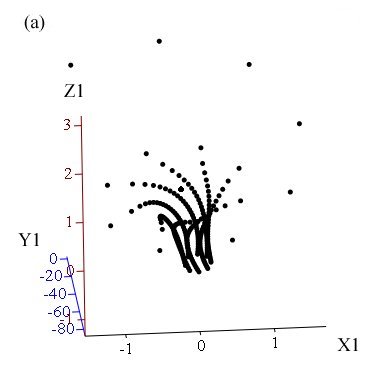}\includegraphics[width=0.5\textwidth]{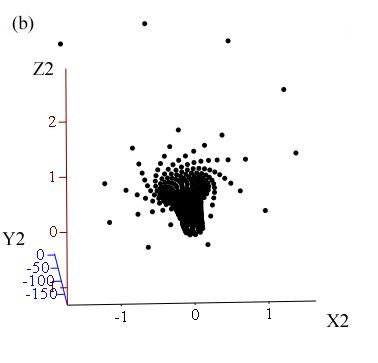}
\includegraphics[width=0.5\textwidth]{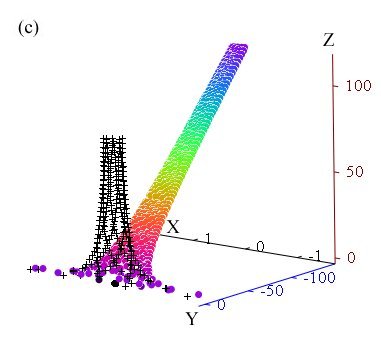}
\caption{The magnetic sub-clouds (a) MC2a, from $t_4$ to $t_4'$, and
(b) MC2b, from $t_4'$ to $t_5$, each of these is modeled as
cylindrical-shape flux tubes, and plotted in the own reference
system (Y-direction represents the cylinder axis). (c) MC2a and MC2b
are plotted in the spacecraft reference system, revealing their
relative positions and encountering angle, after rotation at the
spacecraft hit.} \label{mva2ab}
\end{center}
\end{figure}


By correlating the new time limits resulted from the present study
with the analysis from Article~I we infer new interesting aspects.
Thus, between $t_4$ and $t_4'$, we remark peaks for the rates of
O$^{7+}$/O$^{6+}$, C$^{6+}$/C$^{5+}$ and Fe/O, while for the
$Q_{\mathrm{Fe}}$ a depletion is registered (Figure~3 from
Article~I). Each of these values is higher than the corresponding
threshold, indicating the presence of a MC. Figure~12 from Article~I
reveals the first sector of the MC2 displaying a perturbed helicity
-- this fact could be explained in the light of the new analysis
from this paper. The different angles at which the two clouds hit
the satellite could be an explanation for the strahl asymmetry, but
to have a comprehensive insight into the phenomenon we have pursued
with computations similar to Article~I. We have divided MC2 in MC2a
(from $t_4$ to $t_4'$) and MC2b (from $t_4'$ to $t_5$), and computed
the magnetic field lines using the cylinder-shaped flux rope model
described by \inlinecite{ma07} and adapted for \emph{Ulysses} in
Article~I. This result supports the existence of two distinct
regions in MC2 (Figure~\ref{mva2ab}): MC2a and MC2b. The helicity of
MC2a is right-handed (Figure~\ref{mva2ab}a), while MC2b has a
left-handed helicity (Figure~\ref{mva2ab}b), similar to MC1 and to
the solar source. Figure~\ref{mva2ab}c displays the two clouds
hitting the spacecraft at different angles, with a large tilt angle
between their axes. The situation that MC2a belongs to the same
solar source is difficult to envision. The different helicity of the
two clouds composing MC2 suggests that we have here the case of
merged clouds coming from different solar sources. Most probably
MC2a was produced by a coronal mass ejection (CME) that occurred
behind the limb at a moment close to the CME that produced MC2b. It
seems that MC2b overtook another CME as it travelled. In this way,
two different CMEs arrived and merged at \emph{Ulysses} at different
angles. Their superposition at the beginning of MC2 can explain the
parasite helicity and the intermittency shown by the PADs.


%
\begin{sidewaysfigure}
\centering
\includegraphics[trim=1.5cm 0.0cm 0.5cm 0cm, clip, width=180mm]{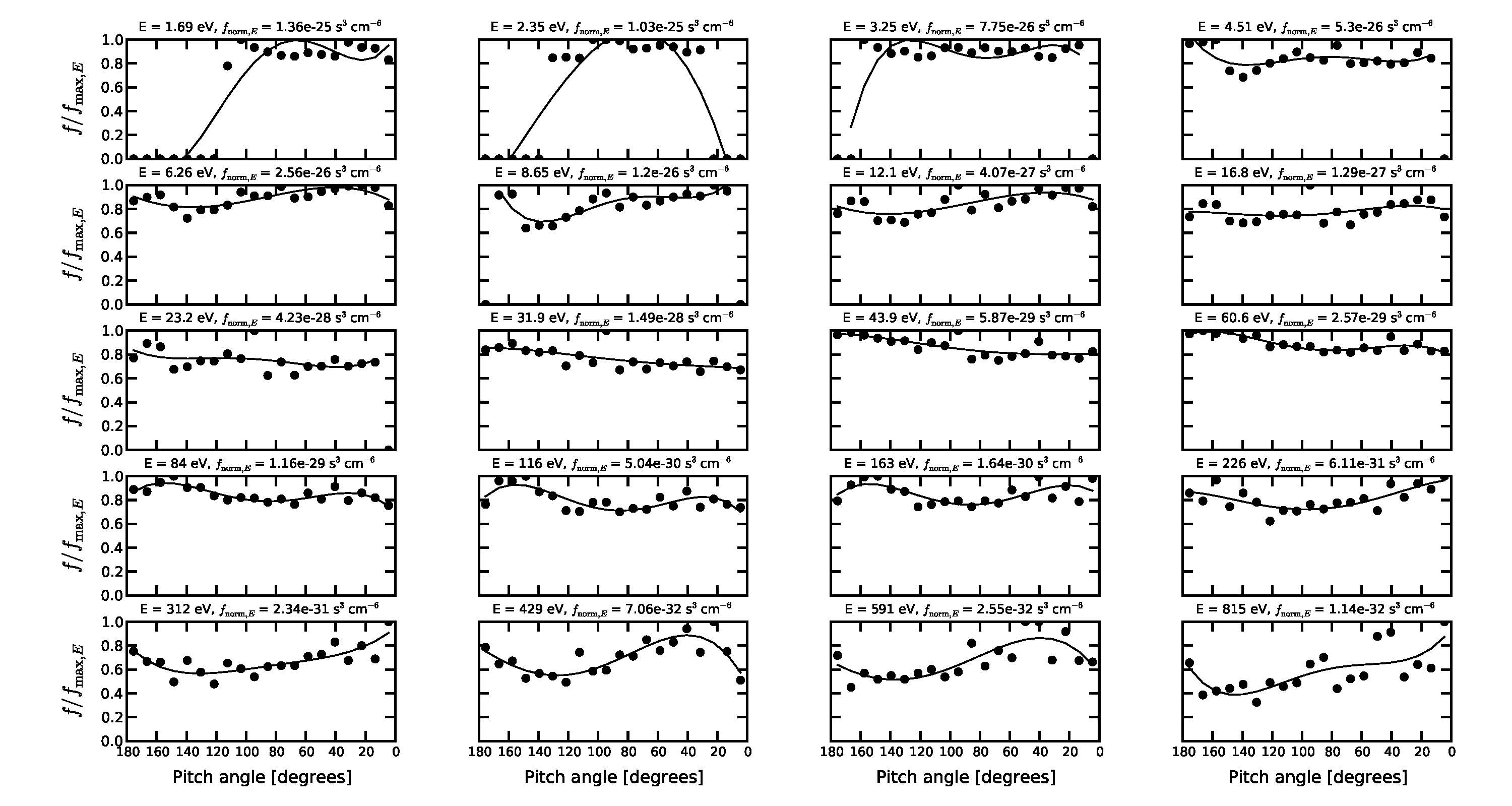}
\caption{PADs for all available energy channels (see Fig. \ref{fb1})
on DOY 19 13:31:41. The first isotropic PAD in the reverse shock.}
\label{fb4}
\end{sidewaysfigure}

After DOY 19 13:31:41 the PADs are isotropized and do not show any
peaks. Figure~\ref{fb4} presents the first snapshot of isotropic
PADs after the CME event. This moment nearly coincides with the time
$t_6 = 19.583 =$ DOY~19 13:59:31 identified in Article~I when
\emph{Ulysses} encounters the reverse shock of the CME. In this
interval the plasma seems to be more turbulent and dominated by
fluctuations of both the particle density and the magnetic field
(see Figure~2 in Article~I). Similarly to the forward shock, the
PADs change sharply after the reverse shock front ($t \geqslant
t_6$), showing rapid and consistent broadening of the suprathermal
population. However, shortly after the shock front the signatures of
an intense antiparallel strahl component, typical for the fast solar
wind, become again visible. In the PADs this strahl appears less
narrow but sufficiently intense to deviate towards lower energy core
populations.

\subsection{CME Precursors}

The arrival time of the forward shock was found in Article~I to be
$t_1=17.875$ (DOY 17 21:00:00), and coincides in Figure~\ref{f2}
(first solid line) with a drastic change of the PADs, namely, an
isotropization of the less intense strahls observed before the
shock. Before this impact the anisotropy of the velocity
distributions presents signatures that indicate the approach of the
CME. These precursors consist of fluxes of streaming and
counter-streaming electrons that are progressively enhanced and
exhibit characteristics similar to those observed later in the
magnetic clouds. The onset of these bi-directional beams prior to
the forward shock suggests the presence of closed magnetic-field
lines propagating ahead of the MC \cite{st11}.

\begin{figure}[h!]
\includegraphics[trim=1.5cm 1.5cm 2cm 1.5cm, clip, width=12cm]{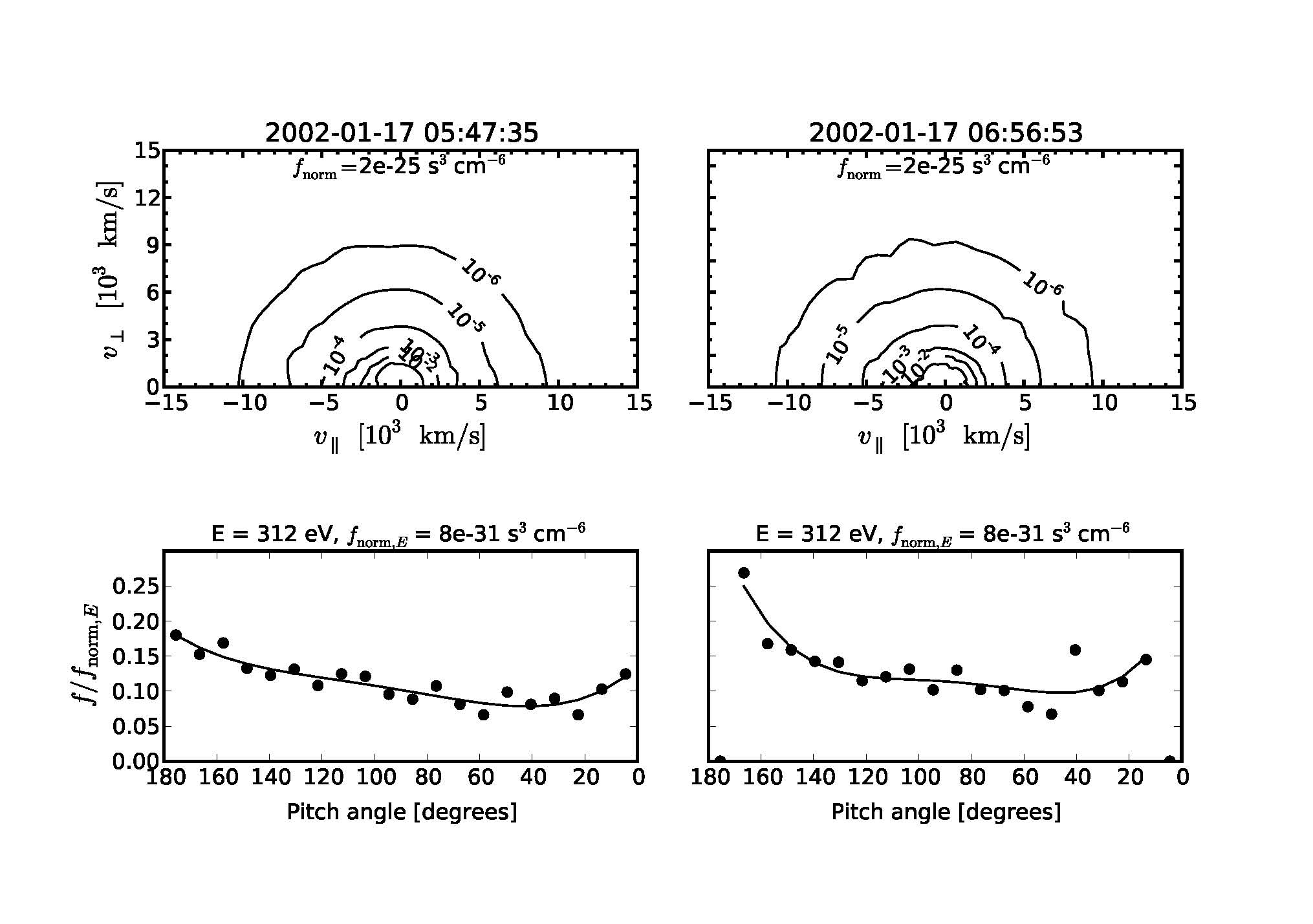}\\
\includegraphics[trim=1.5cm 1.5cm 2cm 1.5cm, clip, width=12cm]{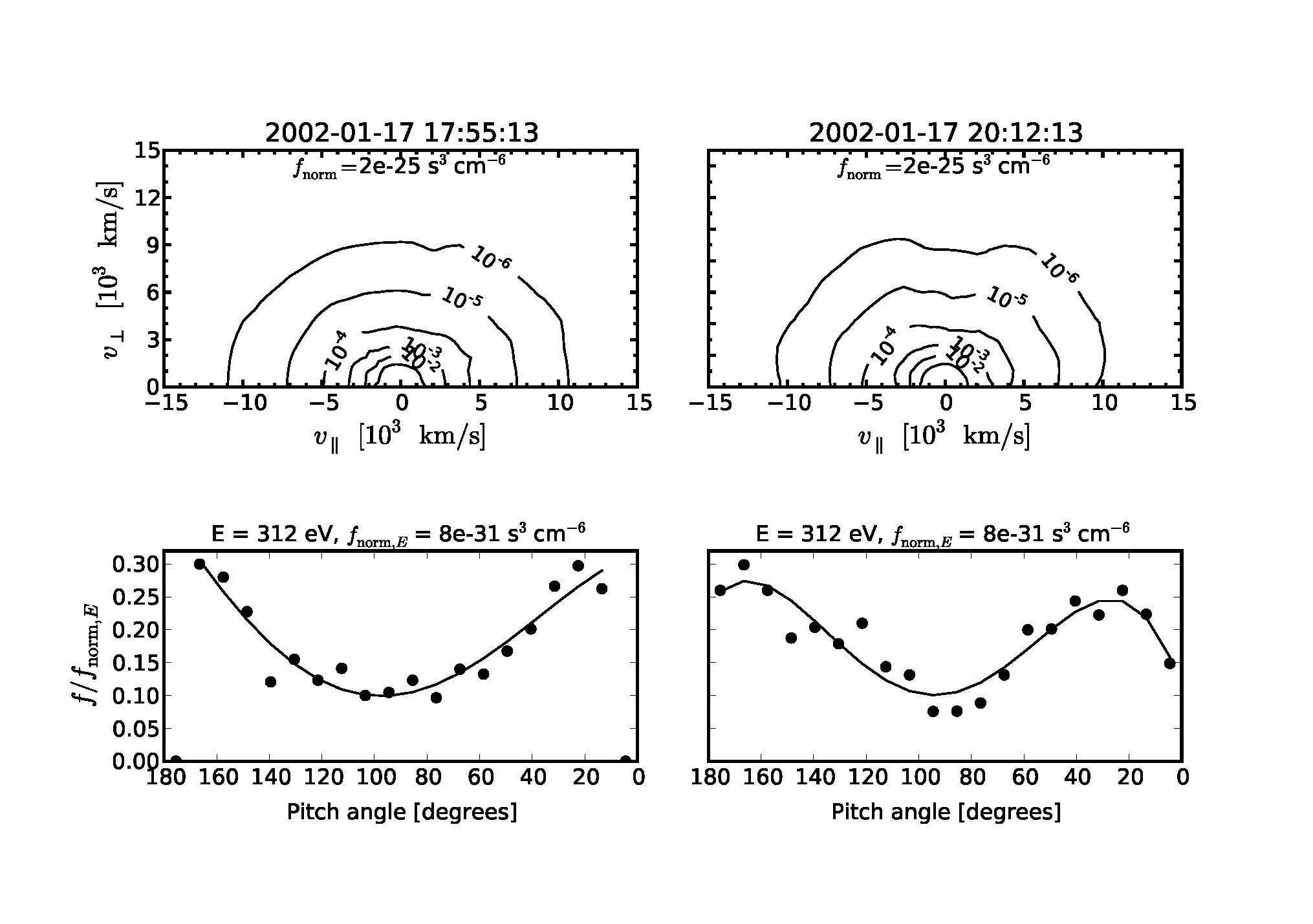}
\caption{Snapshots of velocity distributions (contours) and PADs
($312\;$eV) before the forward shock, showing asymmetric
counter-streams as precursors of the CME counter-streams.}
\label{f3}
\end{figure}

Precursors appear in the PADs of DOY 17 after 05:30:00, see
Figure~\ref{f2}, and rise progressively in two distinct phases. Four
selections of PADs exhibiting signatures of strahls are displayed in
Figure \ref{f3}. As in the CME, these strahls are indicated by peaks
in directions parallel and anti-parallel to the magnetic field.
First, only enhanced anti-parallel (uni-directional) strahls are
observed (top panels) but after DOY 17 09:00:00 (dashed line in
Figure~\ref{f2}), the PADs become more anisotropic showing specific
counter-streaming strahls that are detected first at higher
energies. While the parallel peak is distinguishable only at
sufficiently high energies ($\geqslant 226\;$eV), the anti-parallel
peak continues to be more intense, and is visible at very low
energies down to 8.65 eV. This asymmetry can result, as explained
above, from the superposition of a uni-directional strahl flowing
outward from the Sun (in the fast wind) and the counter-streams
along the closed magnetic-field lines \cite{sk00c}.

The magnetic-field magnitude is still low, and its direction,
reversed after DOY 16 22:00:00 (not shown here), points
predominantly outward from the Sun [$B_{r,t} > 0$] until
\emph{Ulysses} encounters the CME. In this case the more intense
anti-parallel strahl is an inward flow, while the parallel strahl
signifies an outward flow of electrons. All of the magnetic-field
components exhibit a sinusoidal-like variation (see Figure~\ref{f2})
indicating a helical flux rope, but they all keep to small values
and change evenly sign for a few short times, suggesting a magnetic
field topology similar to that at the apex of a loop. Such a
topology of the magnetic field can be responsible for the occurrence
of the counterstreams in the measured distribution.

Shortly after DOY 17 12:00:00, signatures of the parallel strahl
move to lower energies, and counter-streaming strahls appear to be
enhanced. Depletions of the PAD extends to a wide band of energy
from suprathermal energies to less than $60.6\;$eV
(Figure~\ref{f2}). Apparently, the parallel strahl is enhanced at
the expense of the anti-parallel beam. This moment is also followed
by perturbations and broadening of the strahls until DOY 17
17:30:00. The observed broadening of the anti-parallel strahl can be
a result of the pitch-angle scattering in the field fluctuations
driven by the beam-plasma instabilities.

A second enhancement of the counter-streaming strahls is apparent
after DOY 17 15:30:00 (dotted line in Figure~\ref{f2}), and it is
confirmed by the more symmetric and pronounced peaks of PADs, as in
Figure~\ref{f3} (bottom panels). This is a second indication of the
approaching CME, by signatures of a closed magnetic field extending
well ahead of the shock front. These signatures last until $t_1 =$
DOY 17 21:00:00, which is exactly the time estimated in Article~I
for the impact with the forward shock (first solid line in
Figure~\ref{f2}). The last two events displayed in Figure~\ref{f3}
show clear structures (peaks) of counter-streaming strahls as
typical precursors of the CME. We should note that peaks are not
always centered at 0 or 180$^{\rm o}$ pitch-angles. In the right
panels, both peaks deviate from these positions, suggesting the
existence of two pairs of counter-streams (because the distribution
is assumed gyrotropic). These counter-streams represent not only
precursors that could help us to anticipate the upcoming CME, but
their existence also suggests a gradient (or a gradual dissipation)
of the magnetic flux ropes beyond the ICME boundaries.


\begin{figure}[h!]
\includegraphics[width=120mm]{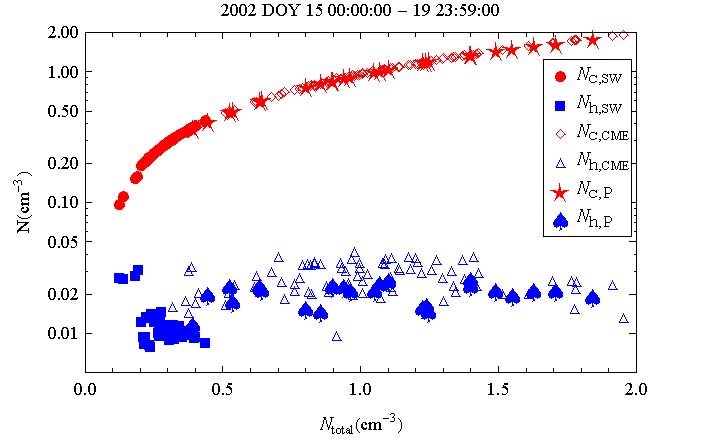}\\
\includegraphics[width=120mm]{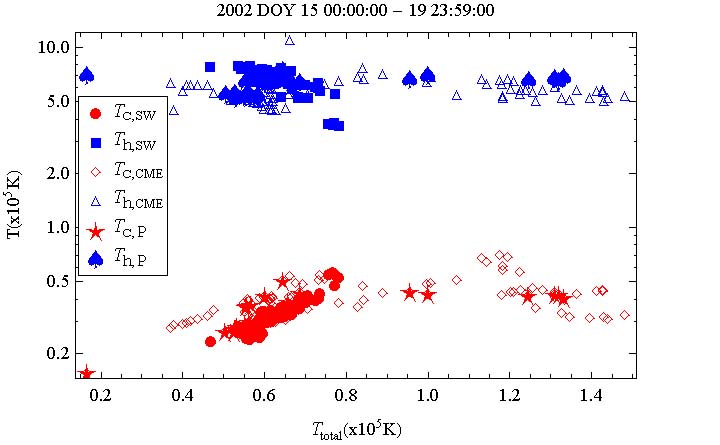}
    \caption{The electron number density (top panel) and temperature
    (bottom panel) for the core (index 'c') and halo
    ('h') components during the CME event (index CME), precursors
    ('P'), and the solar wind ('SW') before the CME.} \label{f10}
\end{figure}

Likewise, indications of the upcoming CME can also be detected in
the solar-wind electron plasma parameters. Figure~\ref{f10} displays
the electron number density (top panel) and the electron temperature
(bottom panel) for the core and halo components during three
distinct intervals of time, namely, the CME event from DOY 18
01:42:00 to DOY 19 23:59:00, precursors in the interval DOY 17
08:00:00--23:59:00, and the solar wind before the CME between DOY 15
02:00:00 -- DOY 18 23:59:00. Additional explanations of these plasma
moments are given in Section 2 above. Distinction can be easily made
since the core is much denser but much cooler than the halo. In the
normal solar wind, unperturbed by the CME, both the core (filled
circles) and halo (filled squares) data concentrate at low values of
density and temperature. The density or temperature contrasts
between these two components are about one order of magnitude on
average. By comparison to the normal solar wind, the CME and
precursors data show the same increase of density and temperature
for both the core and halo populations. Noticeable is the increase
of the core density (top panel) up to one order of magnitude during
the precursors (stars) and the CME (diamonds). The apport of
material is also prominent in the halo component which becomes two
to three times denser in the precursors (filled spades) and CME
(triangles). However, the surplus of kinetic energy is distributed
mainly to the core populations (bottom panel), where the temperature
starts increasing in the precursors (stars) and indicates a peak
during the CME event (diamonds).

During the period of precursor streams the $\mathrm{Fe}/\mathrm{O}$
abundance and the charge-state values of $\mathrm{Fe}$ exhibit a
sudden increase (DOY 17 05:30:30), reaching values of $Q_{\rm
Fe}\approx 14$ that are comparable with the CME (see Article~I,
Figure~3). Also notice that a peak of the charge-state ratio
$\mathrm{C}^{6+}/\mathrm{C}^{5+}$ appears at $t= 17.646=$ DOY~17
15:30:00 (Figure~3 from Article~I) concomitant with a second
enhancement of the precursor counter-streams. During the interval
DOY 17 05:30:30--21:00:00 the C$^{6+}$/C$^{5+}$ ratio can reach
values greater than those observed during the passage of MC2 (see
Article~I, Figure~3), and the solar wind elemental and ionic
composition show an enhancing tendency similar to the upcoming CME.
In addition, the proton density plotted in Figure 1, exhibits a jump
at $t=17.375 =$ DOY 17 09:00:00, 12 hours ahead of the shock
arrival. According to \inlinecite{de13}, such dense region can be
the result of a pressure wave of solar wind material accumulated
from the ambient solar wind ahead of the CME that travels outward
from the Sun in the interplanetary space.

\begin{sidewaysfigure}
\centering
\includegraphics[trim=1.5cm 0.0cm 0.5cm 0cm, clip, width=180mm]{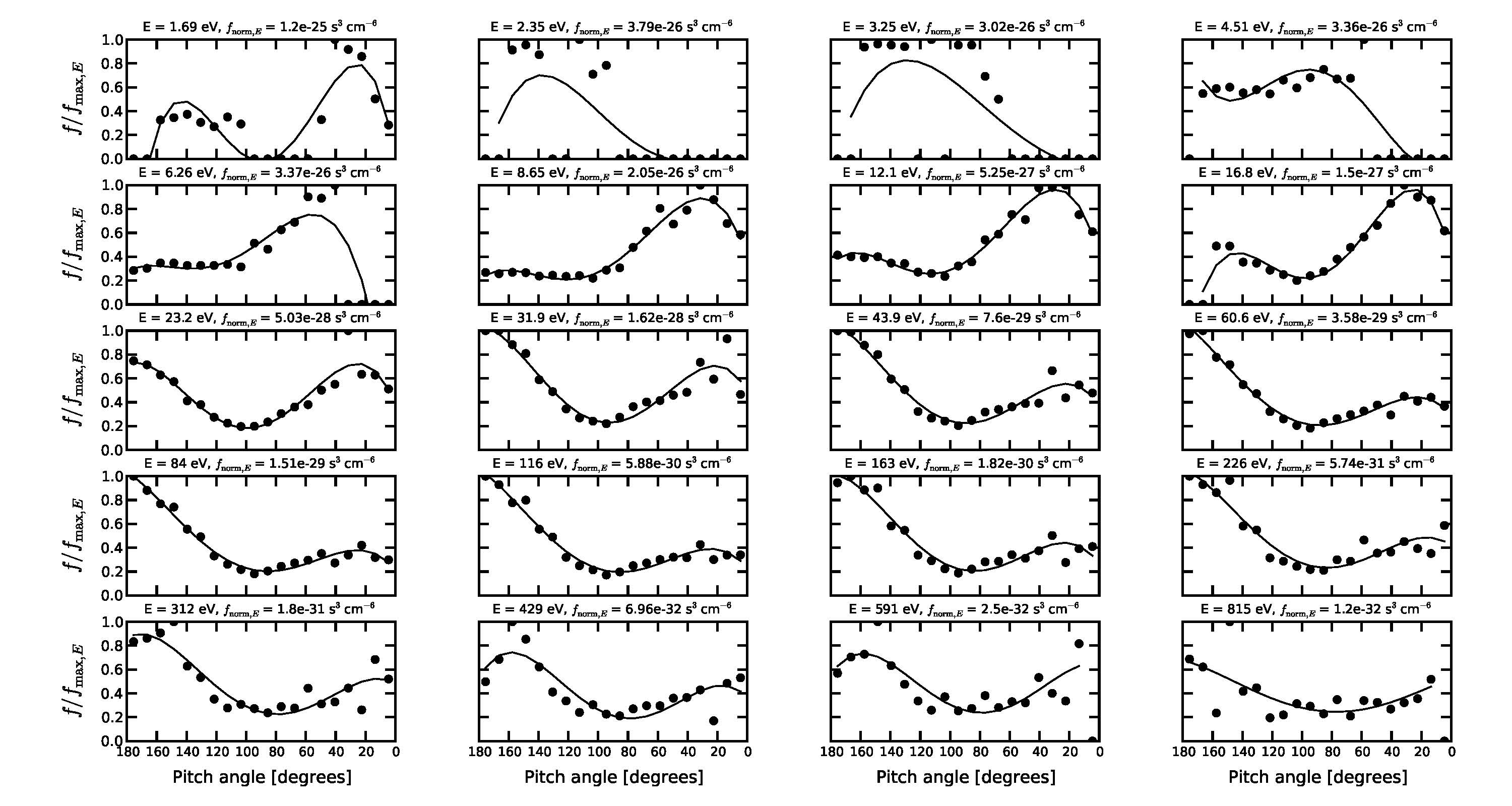}
\caption{PADs for all available energy channels (see Fig. \ref{fb1})
on DOY 14 03:50:15. Nonperturbed (fast) solar wind before the CME
event.} \label{fb5}
\end{sidewaysfigure}
%

\subsection{Strahls in the Fast Solar Wind -- Before the
CME}\label{SWstrahls}

According to observational studies at smaller heliographic distances
and lower latitudes \cite{ma05, an12} signatures of asymmetric or
uni-directional strahls are, in general, present or more pronounced
in the fast wind. Over the time interval examined here,
\emph{Ulysses} was situated at high heliographic latitudes over
coronal holes of the northern pole, and episodes of fast solar wind
are therefore expected to be frequent, except during the CME
intervals.

\begin{figure}[h!]
\includegraphics[trim=1.5cm 0 3.5cm 1cm, clip, width=120mm]{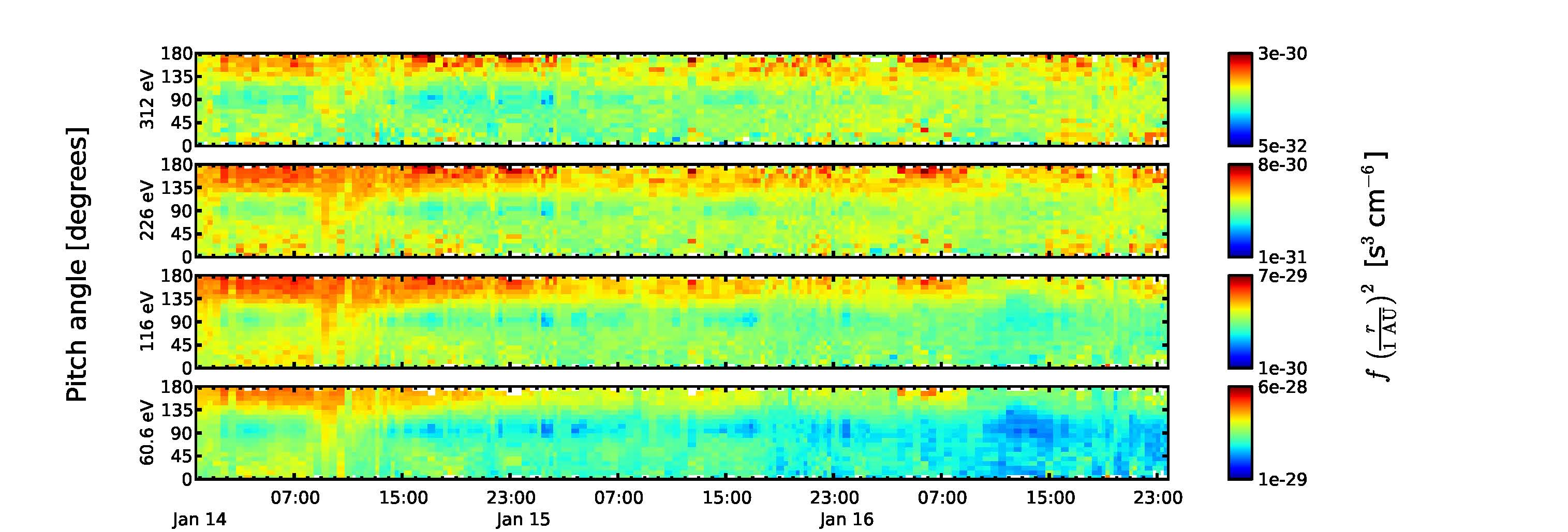}
    \caption{PADs from the fast wind during DOY 14\,--\,16, before the CME event.} \label{f8}
\end{figure}
\begin{figure}[h!]
\includegraphics[trim=1.5cm 1.5cm 2cm 1.5cm, clip, width=120mm]{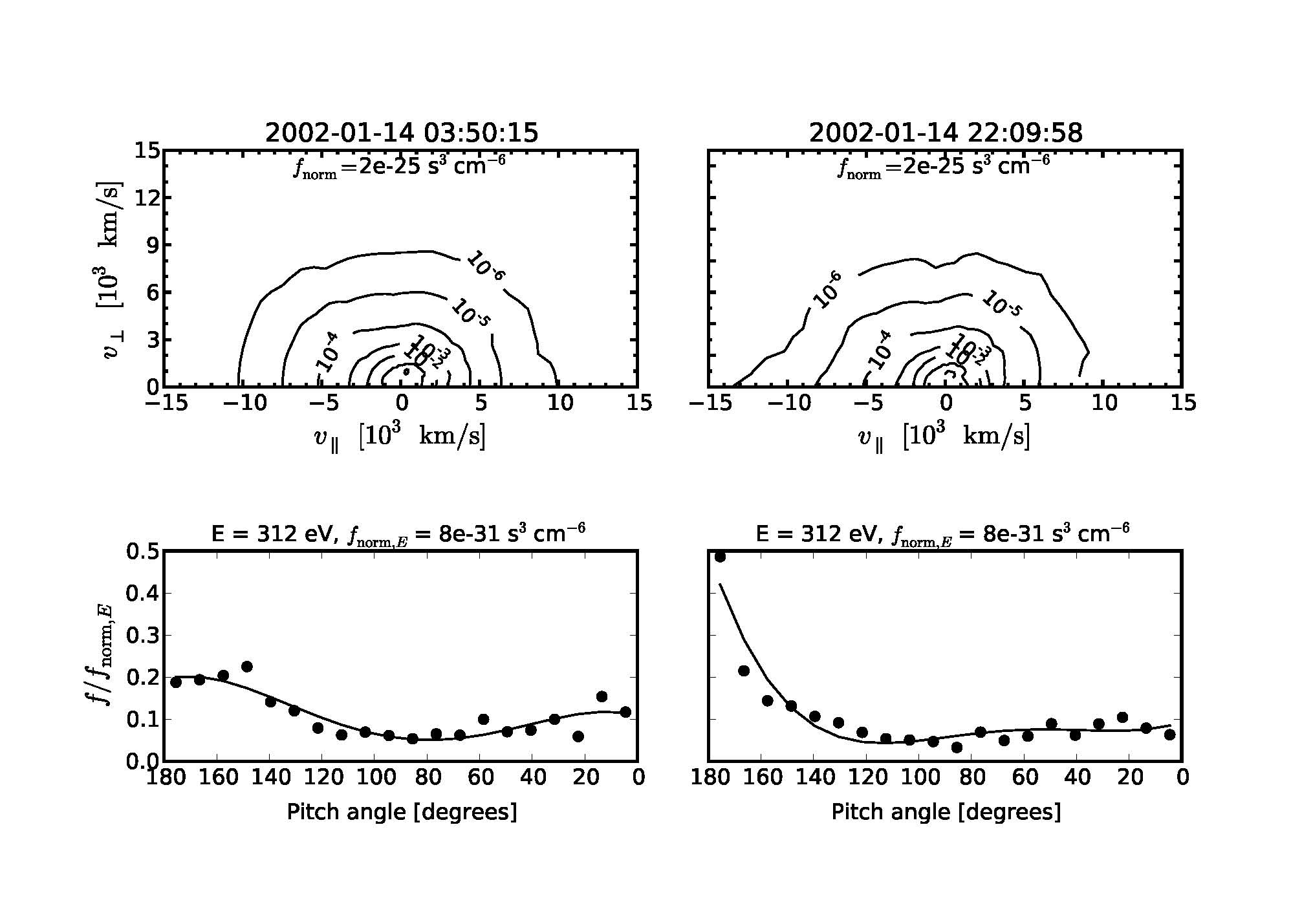}\\
\includegraphics[trim=1.5cm 1.5cm 2cm 1.5cm, clip, width=120mm]{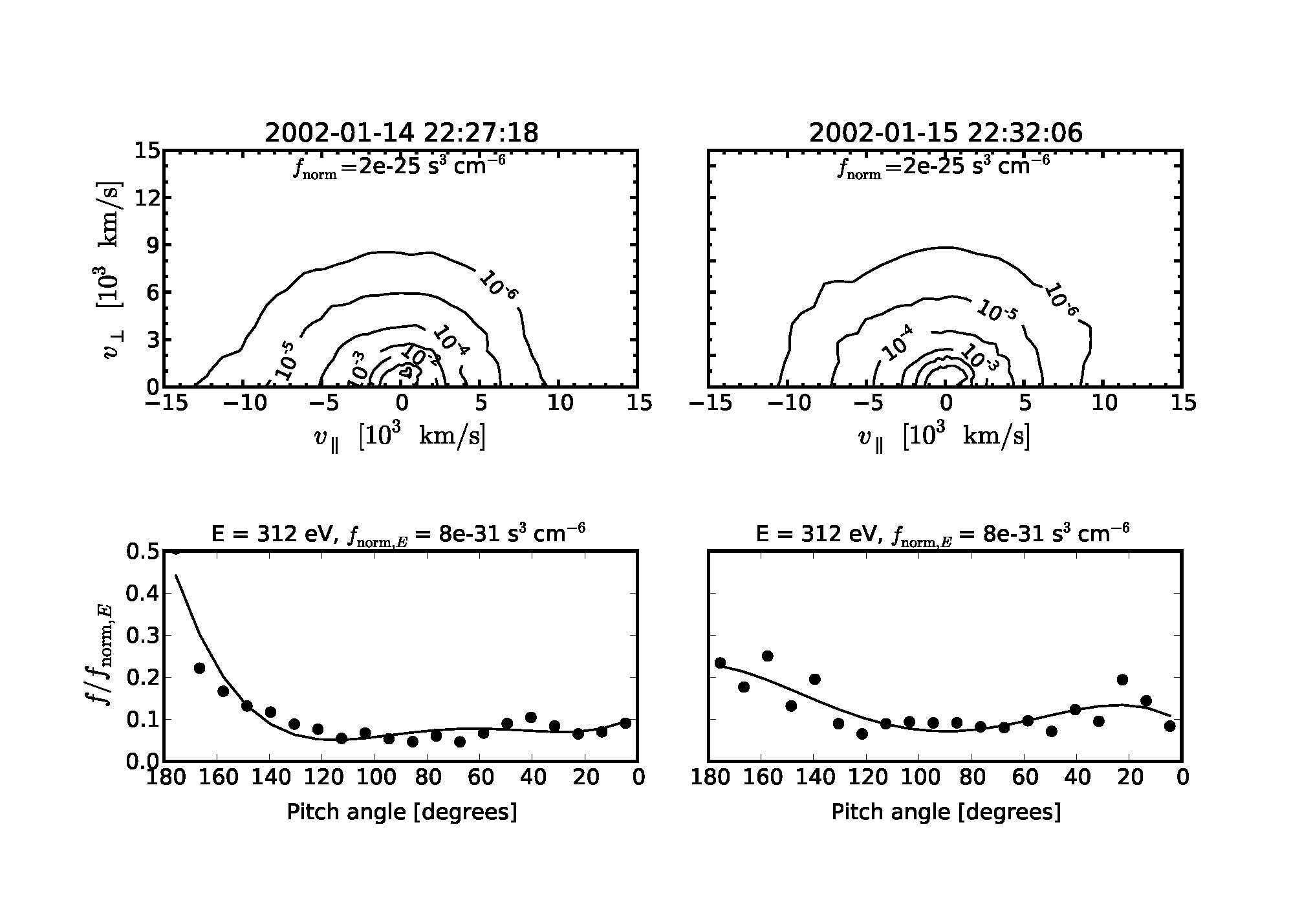}
    \caption{Snapshots of velocity distributions (contours) and PADs ($312\;$eV) from
    the fast wind on DOY 14\,--\,15 before the CME.} \label{f9}
\end{figure}

The  long-lasting fast wind episode between days 7 and 16 (see
Figure~\ref{f1}) is particularly important for a complete overview
of our event. Although the solar wind is fast and very fast
(reaching peaks $\gtrsim 800\;$km/s), this is a quiet time, without
noticeable fluctuations of the magnetic field and density. The
radial component of the magnetic field indicates a constant inward
orientation ($B_r <0$) for the whole interval of ten days, with a
short exception between DOY 16 04:00:00\,--\,09:00:00, in the day
before the CME precursors. PADs are plotted in Figure~\ref{f8} for
the last three days of this interval, showing a prominent
anti-parallel peak of intensity lower than that observed in the
precursors and CME. (The comparison is possible since we have used
the same color bars as in Figure~\ref{f2}). In this case the
antiparallel peak indicates an outward moving flow of electrons, the
presence of which is a characteristic of the whole interval of fast
wind.

Less intense parallel peaks are also visible at suprathermal
energies during DOY 14. An examination of the full energy map in
Figure~\ref{fb5} (Appendix~B) reveals an enhancement of this
parallel peak at lower energies, where it becomes more prominent
than the anti-parallel strahl. The asymmetry of these peaks seems to
be similar to what we found at the entrance into the CME and
supports the existence of strahls and not just a simple effect of
the (parallel) temperature anisotropy (that should produce more
symmetric peaks). This asymmetry is reduced and PADs appear less
anisotropic in the last day (DOY 16) of this interval, probably as a
side effect of the continuous drop of energetic particles. Such
intervals of a low-energetic particle intensity can be attributed to
merged interaction regions (MIRs) \cite{ri02}, but the relatively
low magnetic-field strengths (Figure~\ref{f1}) and low
$\alpha$-particle densities (not shown here) do not sustain the
existence of a MIR in this interval.

It is difficult to select an event (or several events) with PAD
profiles to be representative of the whole interval, even for these
conditions of a less perturbed solar wind. However, we have
identified a restrained variety of strahls typical for this
interval, and such cases are illustrated in Figure~\ref{f9} using
the same plots of the velocity distributions (isocontours) and PADs
(only for 312 eV). The top-left panels (DOY 14 03:50:15) present
signatures of two opposite, parallel and antiparallel strahls, with
similar intensities but different angular widths. The next event on
DOY 14 22:09:58 exhibits a single, but more intense antiparallel
strahl. Immediately after (DOY 14 22:27:18, bottom-right panels)
counterstreams rise with different intensities but similar angular
widths. The last case in the bottom-left panels show a distribution
less anisotropic in the parallel direction but with a number of
secondary beams at oblique directions.

\section{Summary and Conclusion}

This article presents a qualitative analysis of the electron beaming
strahls in the solar wind during a high-latitude CME registered by
\emph{Ulysses} in Jannuary 18\,--\,19, 2002. As a distinct component
of energetic particles, the strahls are expected to be more
pronounced in the observations from high-latitude and from the
passage of a CME. We have described the evolution of the velocity
distributions in the magnetic clouds, ahead of the forward shock and
after the reverse shock. Uni-directional and bi-directional strahls
were identified as peaks of intense fluxes of electrons, which are
in general directed parallel or anti-parallel to the magnetic field
direction. To objectively assess the presence and extent of the
strahls, their signatures were inspected at all energy channels
available from observations.

Difficulties arise in the analysis of distributions with a symmetric
depletion or two symmetric peaks, which can be the result of either
two counter-streaming strahls or just an excess of parallel
temperature. For these cases we have shown that a distinction is
possible by an extended comparison of the PAD profiles at different
(lower) energy channels. The symmetry of two opposite strahls breaks
down at some energy levels, showing different profiles with clear
signatures of asymmetric or uni-directional peaks. These differences
are expected to appear since the parallel and anti-parallel strahls
have different paths and origins. A plausible explanation for the
asymmetry of the counter-streaming strahls (\emph{e.g.} intensities
and angular widths in Figures~\ref{f6} and \ref{f7}) can be given by
the so-called strahl-on-strahl model, originally proposed by
\inlinecite{sk00c}. The strahl-on-strahl distributions can result
from the combination of electrons on their first transit outward
from the Sun (narrow beam from the fast wind) with populations
trapped on field lines still connected to the Sun at both ends
(broad, bi-directional beams). Thus, these distributions and their
evolution provide information about the magnetic topology of the
solar wind and its variations across an CME.

A complete scan of the PADs has also revealed that, whether these
strahls are uni- or bidirectional, their signatures can extend to
very low energies, suggesting that their electron densities become
in this case comparable with that of the thermal-core populations.
This feature is not only a characteristic of the CME distributions,
when the strahls are indeed enhanced, but it can be easily observed
in the fast wind before the CME. It can also be correlated with a
drop of the break point with increasing radial distance from the
Sun.

%
%
The temporal evolution of the PADs shows two large intervals of
continuous counter-streaming patterns during the CME event. At
sufficiently high energies, \emph{e.g.} 226 or 312 eV, the presence
of counter-streams exceeds 85 \% of the CME duration, estimated in
Article~I between the forward and reverse shocks, \emph{i.e.}
$[t_2,~t_6]$. With small adjustments of one to two hours these two
boundaries are confirmed by drastic changes of the PADs, namely, the
first indications of strahls at $t_2$, and then their isotropization
at $t_6$. If we look to the other two internal boundaries [$t_3$ and
$t_4$], attributed in Article~I to the end of MC1 and the beginning
of MC2, respectively, the first one does not indicate any major
change of PAD, but the second one ($t_4$) marks the end of a long
interval (ten hours) of continuous presence of the counter-streaming
strahls. This interval starts at $t_2'$, which is a new internal
boundary indicated by the limits of intense counter-streams.

After $t_4$ the anti-parallel strahl is constantly present, while
the parallel strahls appear only sporadically until $t_4'$, when the
counter-streams become continuous again. Similarities of these two
intervals $[t_4,~t_4']$ and $[t_2,~t_2']$ support the existence of
two magnetic clouds suggested by the results of Article~I. The
counter-streams can be observed until $t_6$, but shortly after $t_5$
the anti-parallel strahl becomes markedly less intense than the
parallel strahl. We have in this case a very clear bipolar
signature, which starts with an intense anti-parallel strahl between
$[t_2,~t_2']$, and ends with a more enhanced parallel strahl in the
interval $[t_5,~t_6]$. This is confirmed by the large-scale trend of
the magnetic-field components [$B_{r,n}$], which rotate from
negative to positive values in the CME interval $[t_2,~t_6]$.
Correlating with the analysis in Article~I, the new internal
boundaries can be associated with additional local jumps and
depletions in the temporal evolution of the elemental composition
and charge states, \emph{e.g.} peaks of the rates of
O$^{7+}$/O$^{6+}$, C$^{6+}$/C$^{5+}$ and Fe/O are observed in the
interval $[t_4, ~t_4']$, but these are not reproduced in the
interval $[t_2, ~t_2']$.

In the second part of our analysis, we have identified
counter-streaming electrons which appear well in advance of the
forward shock, as precursor elements -- less intense but very
similar to the counter-streams in the CME. Some enhanced traces of
uni-directional strahls appear first, before the counter-streams,
but these are ubiquitous in the fast wind over the poles, and
probably cannot be attributed exclusively to an approaching CME. The
first counter-streaming signatures are transmitted by the
suprathermal channels (\emph{e.g.} 226 and $312\;$eV), with more
than five hours in advance of the shock front. This interval is much
longer than that of only one hour reported by ACE for the onset of
the bi-directional beams prior to a MC in the ecliptic \cite{st11}.
The CME expansion with heliocentric distance and heliographic
latitude could explain this difference, if we admit the existence of
bipolar magnetic-field lines ahead the MC, and that these fields
undergo the same effect of expansion. Additional confirmations for
the existence of these precursors, including their temporal extent,
were obtained by testing the solar-wind electron-plasma parameters
and the elemental abundance. By comparison to the normal solar wind,
the CME and precursor data show the same enhanced values of electron
density and temperature, for both the core and halo populations.
Accumulations are also detected in the charge-state composition such
as O, C, and Fe, as reliable signatures of an CME.

In the unperturbed solar wind the most visible is a uni-directional,
anti-parallel strahl with densities one or two orders of magnitudes
less than the thermal core populations. Streaming (and sometimes
counter-streaming) patterns in the solar wind are markedly different
from those observed during the CME and precursors. Visual comparison
of the peak widths and intensities confirms the expectation that
strahls are markedly enhanced by the apport of coronal matter from
CME.

To conclude, the electron distributions are found to be highly
anisotropic and dominated by the presence of strahls, not only
during the CME but also well before the forward shock. Evidence
presented here strongly supports a major implication of these
populations into the interplanetary manifestation of the CME. The
boundaries estimated in Article~I for the distinct structures of the
CME, \emph{i.e.} shocks and magnetic clouds, are reconfirmed here.
With only one exception, these limits correspond to significant
changes of PADs (see Figure~\ref{f2}). The exception is given by the
ending time of MC1 (as estimated in Article~I), which here indicates
only a slight remission of intensity of the parallel strahl.
Likewise, the analysis of uni-directional or bi-directional strahls
enabled us to identify CME precursors before the forward shock, as
well as substructures of the magnetic clouds.

\begin{acks}
The authors acknowledge use of the \emph{Ulysses}/SWOOPS electron
pitch-angle data from the ESA-RSSD Web service:
{www.rssd.esa.int/pub/ulysses/}, PIs: R. Skoug and D.J. McComas; and
CDAWeb service: {cdaweb.gsfc.nasa.gov}, data provider, A. Balogh
(magnetic field), J.L. Phillips (plasma). ML acknowledges financial
support from the EU Commission and Research Foundation Flanders
(FWO) as FWO Pegasus Marie Curie Fellow (grant 1.2.070.13) and
partial support from the Ruhr-Universit\"at Bochum, and the Deutsche
Forschungsgemeinschaft (DFG), grant Sh 21/3-2. These results were
obtained in the framework of the projects GOA/2009-009 (KU Leuven),
G.0729.11 (FWO-Vlaanderen) and C~90347 (ESA Prodex 9). The research
leading to these results has also received funding from the European
Commission's Seventh Framework Programme (FP7/2007-2013) under the
grant agreements SOLSPANET (project nÂ° 269299,
{www.solspanet.eu}), SPACECAST (project nÂ° 262468,
{fp7-spacecast.eu}), eHeroes (project nÂ° 284461,
{www.eheroes.eu}) and SWIFF (project nÂ° 263340, {www.swiff.eu}).
The authors are grateful to Horst Fichtner for reading the
manuscript and providing insightful remarks, and the anonymous
reviewer for his/her extremely thoughtful and constructive comments
on the manuscript.
\end{acks}



\end{article}
\end{document}